\documentclass[aps,floats,twocolumn,epsf,prb,showpacs]{revtex4-1}
\usepackage{graphicx}
\usepackage{epstopdf}
\usepackage{hyperref}
\hypersetup{colorlinks=true,urlcolor=blue,citecolor=blue,linkcolor=blue,breaklinks=true}
\usepackage{color}
\usepackage{here}
\usepackage{amssymb}   
\usepackage{amsmath}

\newcommand{\Gd}{\widetilde{G}}\newcommand{\Sigmad}{\widetilde{\Sigma}}
\newcommand{\Sigmaloc}{\Sigma_{\mathrm{loc}}}
\newcommand{\SigmaDF}{\Sigma_{\mathrm{DF}}}

\newcommand{\Fd}{\widetilde{F}}
\newcommand{\Floc}{F_{\mathrm{loc}}}
\newcommand{\Gloc}{G_{\mathrm{loc}}}
\newcommand{\Phid}{\widetilde{\Phi}}
\newcommand{\Gammad}{\widetilde{\Gamma}}

\newcommand{\vvw}{{\nu\nu'\omega}}
\newcommand{\kkq}{{kk'q}}
\newcommand{\veck}{\mathbf{k}}
\newcommand{\vecq}{\mathbf{q}}

\newcommand{\Phiph}{\Phi_{ph}}
\newcommand{\Phipht}{\Phi_{\overline{ph}}}
\newcommand{\Phipp}{\Phi_{pp}}

\newcommand{\pht}{\overline{ph}}

\begin{document}

\title{Parquet dual fermion approach for the Falicov-Kimball model}

\author{K. Astleithner}
\author{A. Kauch}
\email{kauch@ifp.tuwien.ac.at}
\author{T. Ribic}
\author{K. Held}

\affiliation{Institute of Solid State Physics, TU Wien, 1040 Vienna, Austria}

\date{ \today }

\begin{abstract}
In the Falicov-Kimball model, a model for (annealed) disorder, we expect  weak localization corrections to the optical conductivity. However, we get such weak localization effects only when employing  a $pp$-ladder approximation in the dual fermion approach. In the full parquet approach these  $pp$-contributions  are suppressed by $ph$-reducible diagrams. For the optical conductivity, we find that the $\pht$-channel yields the main contribution, even in the region where weak localization in the $pp$-ladder was indicated.
\end{abstract}

\maketitle

\section{Introduction}

The Falicov-Kimball model (FKM) \cite{Falicov1969} is one of the simplest models for electronic correlations and describes fully immobile electrons that interact with mobile conduction electrons. In this sense it can be seen as a simplified version of the Hubbard model \cite{Hubbard1963}, where one spin species is assumed to be localized, and hopping is allowed for the other spin species only. Despite its simplicity, finding a solution to the FKM remains  challenging.

For the two dimensional FKM, a phase transition towards a checkerboard charge density wave (CDW) was proven to occur at and close to half-filling \cite{Kennedy1986, Brandt1986}, as well as a metal-to-insulator transition. The FKM can further be solved (semi)-analytically in infinite dimensions using  dynamical mean field theory (DMFT) \cite{Metzner1989, Georges1992a,Janis1991}. This can also be considered as an approximation for a finite dimensional system, where all local correlations are taken into account. {DMFT is also a good approximation for somewhat higher temperatures where the disorder on each site acts uncorrelated  and results in a  temperature-independent solution.} 

Most of the DMFT results in the FKM have been reviewed in a seminal oeuvre by Freericks and Zlati\'{c} \cite{Freericks2003} \footnote{For the extended FKM cf.~Refs.~\onlinecite{Lemanski2017} and \onlinecite{Kapcia2019}}. However, the physics of the FKM is mainly governed by CDW fluctuations, so that  nonlocal correlations play the key role in the paramagnetic phase. To include these in addition to the local ones already fully covered in DMFT, cluster \cite{Maier2005} and diagrammatic extensions of DMFT have been developed. The latter include the dynamical vertex approximation (D$\Gamma$A) \cite{Toschi2007,Katanin2009,Kusunose2006} and the dual fermion (DF) approach  \cite{Rubtsov2008}. In a similar development for disordered systems, Jani\v{s}\cite{Janis2001} developed vertex corrections to the coherent potential approximation (CPA).  For a review see Ref.~\onlinecite{RMPvertex}. As D$\Gamma$A is not easily applied to the FKM, requiring mixed vertex functions of the mobile and immobile electrons, we choose to employ the DF approach for the FKM using both the parquet equations and a ladder approximation, extending earlier approaches using ladder DF calculations only \cite{antipov2014, ribic2016, yang2014}. We compare the full parquet DF approach to the simple ladder approach and also analyze the different contributions from the different channels: particle-hole ($ph$), transversal particle-hole ($\pht$) and particle-particle ($pp$).

Specifically, we investigate the effect of nonlocal correlations as resulting from the DF approach onto the optical conductivity, describing the interaction of the system with light. The FKM is a model describing annealed disorder. And it is known that for disordered systems weak localization \cite{Altshuler1985} (corresponding to diagrams in a $pp$-ladder) leads to a diminution of the DC optical conductivity and therefore an enhancement of the electrical resistivity, even when there is no gap in the one-particle spectrum. { For more recent studies of weak localization in the Falicov-Kimball model, see Refs.~\onlinecite{Antipov2016,Zonda2019}}. We confirm the appearance of weak localization in the FKM via an employment of the $pp$-ladder series. However this effect is 
superseded by the dominating contribution of the $\pht$-channel to the optical conductivity in the full parquet approach.

The outline of the paper is as follows:
In Sec. \ref{sec:model} we introduce the FKM and the properties of the local vertex function in DMFT, which at self-consistency is employed as the basic building block in the DF parquet approach. Then the different methods, the parquet approach and ladder approximations, that are employed as well as the corresponding equations are explained.  In Sec. \ref{sec:results}, numerical results for the self-energy, the optical conductivity and its corresponding current-current correlation function and the charge susceptibility are presented: in Sec. \ref{sec:halffill} for a half-filled FKM, and in Sec. \ref{sec:doped} for a doped system with filling $n_c=0.15$ for the $c$ electrons and $n_f=0.5$ for the $f$ electrons. Our main findings are finally summarized in Sec. \ref{sec:conclusion}.

\section{Model and methods}
\label{sec:model}
\subsubsection*{Falicov-Kimball model}
The Hamiltonian of the one-band (spinless) Falicov-Kimball model (FKM) reads

\begin{eqnarray}
\mathcal{H} &=& -t\sum_{\langle ij\rangle}c^\dagger_ic_j^{\phantom{\dagger}} +U\sum_ic^\dagger_ic_i^{\phantom{\dagger}} f^\dagger_if_i^{\phantom{\dagger}} \nonumber \\
&&-\mu\sum_i(c^\dagger_ic_i^{\phantom{\dagger}} +f_i^\dagger f_i^{\phantom{\dagger}} )+\varepsilon_f\sum_if^\dagger_if_i^{\phantom{\dagger}}  \;.
\end{eqnarray}
Here $c^\dagger_i$ ($c_i$) create (annihilate) a mobile electron and  $f^\dagger_i$ ($f_i$) a localized electron at lattice site $i$. The hopping $t$ between nearest neighbors is allowed only for mobile electrons in the FKM, and the local Coulomb interaction $U$  acts between an itinerant and a localized electron at the same site; $\mu$ and $\varepsilon_f$ denote the chemical potential and the local potentials for the $f$ electrons, respectively. We choose our units of energy as $D\equiv4t\equiv1$, $k_B\equiv1$ and the Planck constant $\hbar\equiv1$. Furthermore, when calculating the optical conductivity, we set the lattice constant $a\equiv1$ and the elementary charge $e\equiv1$.

\subsubsection*{Local two-particle vertex}

\begin{figure}
\begin{center}
\includegraphics[width=0.5\textwidth]{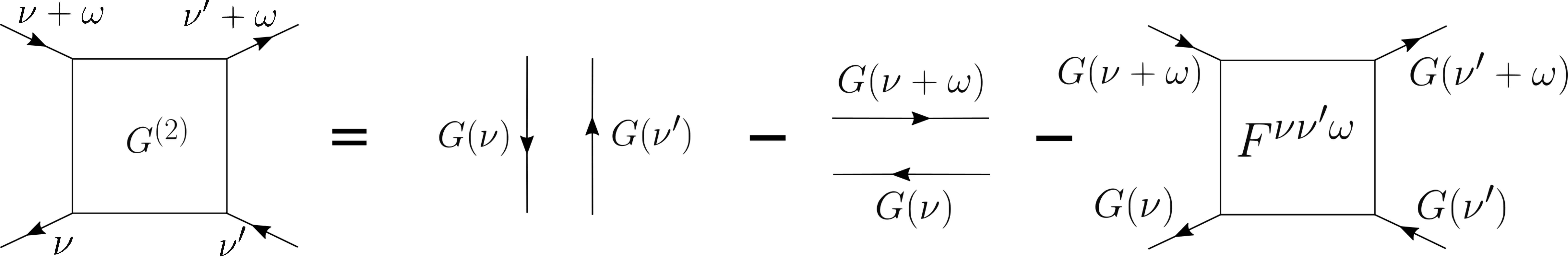}
\caption{{}{The two-particle Green's function $G^{(2)}$ consists of two pairs of disconnected Green's function lines and the connected vertex function $F^\vvw$. In this figure only the frequency arguments are shown.}}
\label{fig:G2}
\end{center}
\end{figure}

The two-particle vertex function $F^{\kkq}$  {}{ is defined as the connected part of the two-particle Green's function $G^{(2)\kkq}$ with the incoming and outgoing lines amputated (also see Fig.~\ref{fig:G2})}:
\begin{equation}
\begin{aligned}
G^{(2)\kkq}&=\beta G(k)G(k')\delta_{q0}-\beta G(k)G(k+q)\delta_{kk'}\\
&-G(k)G(k+q)F^\kkq G(k')G(k'+q).
\label{Gtwop}
\end{aligned}
\end{equation}
Here and in the following, we use a four-vector notation $k=(\veck,\nu)$ and  $q=(\vecq,\omega)$ which subsumes both the momenta and the corresponding Matsubara frequencies.

The local two-particle vertex function $\Floc^{\nu\nu'\omega}$  of the itinerant electrons (i.e. the connected part of the local two-particle Green's function $\Gloc^{(2)\vvw}$) for the two dimensional FKM  can be calculated (semi-)analytically in DMFT. As the mobile electrons can only scatter indirectly via the localized electrons, and are otherwise noninteracting, the local vertex function exhibits a reduced frequency structure, only having finite values for $\omega=0$ and for $\nu=\nu'$. The analytical expression for $\Floc$ can be shown to have the following form \cite{ribic2016}:

\begin{equation}
\Floc^{\nu\nu'\omega}=\beta(\delta_{\omega,0}-\delta_{\nu,\nu'})a(\nu)a(\nu'+\omega),
\label{equ13}
\end{equation}

where $a(\nu)$ is given by

\begin{equation}
a(\nu)=\frac{(\Sigmaloc(\nu)-U)\Sigmaloc(\nu)}{\sqrt{p_1 p_2}U}.
\end{equation}
Here, $\beta=1/T$ denotes the inverse temperature, $\Sigmaloc$ the local DMFT self-energy and $p_2\equiv1-p_1\equiv1-\langle f^\dagger_if_i\rangle$ the number of sites without localized electrons. The number of localized electrons is also referred to as $n_f\equiv p_1$ below, and that of the mobile electrons as $n_c$.

\subsubsection*{Parquet equation}

\begin{figure}
\begin{center}
\includegraphics[width=0.5\textwidth]{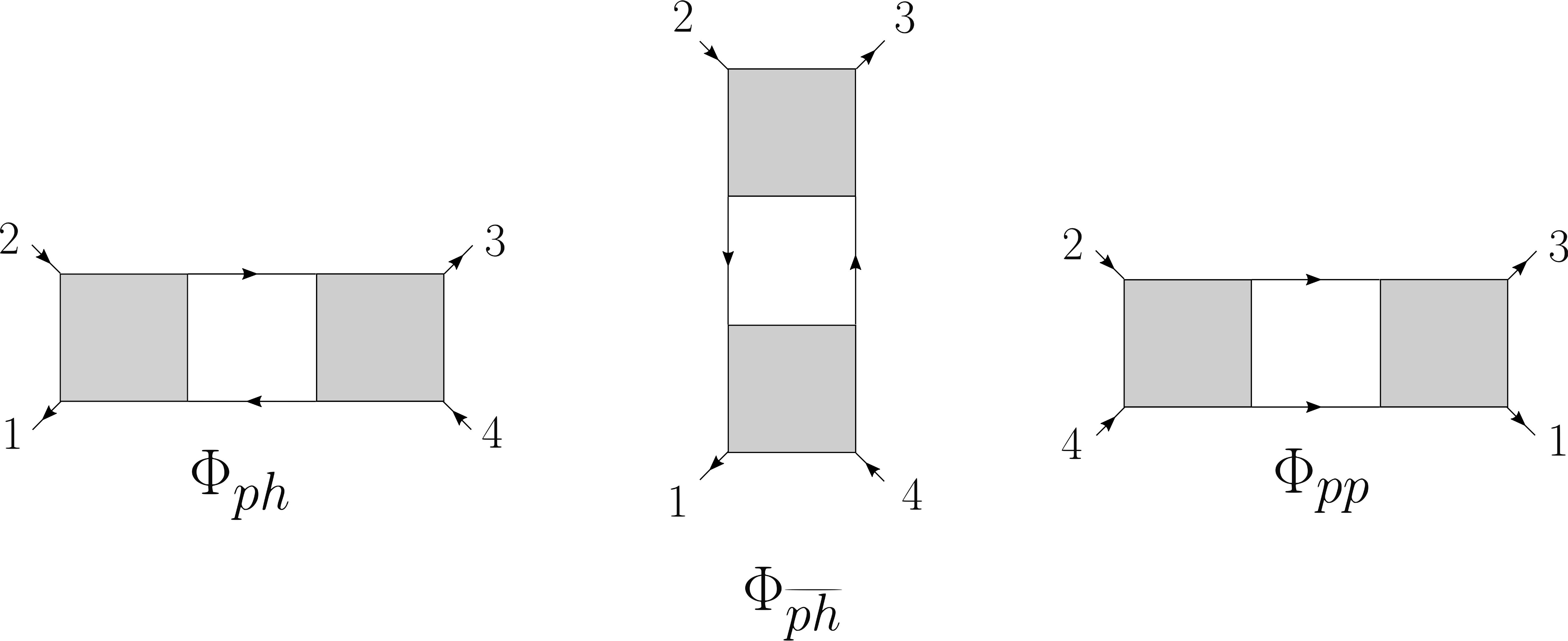}
\caption{{}{A two-particle reducible diagram can be assigned to exactly one of three channels, according to which two of the four outer legs of the diagram can be separated from the other two. In the particle-hole channel $\Phiph$ legs (1,2) are separated from (3,4), in the transversal particle-hole channel $\Phipht$ (2,3) are separated from (1,4) and in the particle-particle channel $\Phipp$ (2,4) are separated from (1,3). }}
\label{fig:red}
\end{center}
\end{figure}

{}{The two-particle vertex $F^{\kkq}$ can be represented by a sum of diagrams that are classified according to their reducibility~\footnote{A diagram is two-particle reducible if it can be separated into two diagrams by cutting two Green's function lines.}. We distinguish four types of diagrams: a class of fully irreducible diagrams (contained in the fully irreducible vertex $\Lambda$) and three classes of reducible diagrams (see Fig.~\ref{fig:red}): (i) diagrams reducible in the particle-hole channel (contained in the reducible vertex $\Phi_{ph}$), (ii) in the particle-hole transversal channel ($\Phi_{\pht}$) and (iii) in the particle-particle channel ($\Phi_{pp}$). The so-called parquet equation than reads:
\begin{equation}
F = \Lambda +  \Phi_{ph} + \Phi_{\pht} + \Phi_{pp}.
\label{eq:pae}
\end{equation}
The reducible vertices $\Phi_r$ in the three channels $r\in\{ph,\pht,pp\}$ can be obtained through the respective Bethe-Salpeter equations. The fully irreducible vertex $\Lambda$ is not given by the parquet approach and has to be provided as input. In the lowest order  $\Lambda$ is equal to the bare Coulomb interaction $U$. Taking $\Lambda=U$ amounts to the so-called parquet approximation.} This kind of channel decomposition is similar (except for the spin) as for the Hubbard model~\cite{Kauch2019}. 

 The FKM describes the $f$-$c$ interaction or annealed disorder. The diagramatics for quenched, uncorrelated disorder is different.

\subsubsection*{Parquet dual fermion approach}
{The DF approach maps an interacting lattice model onto a set of interacting local problems which couple to non-local "dual" fermions.} The local vertex function $\Floc$ {of the local problems} is the basic building block for the DF approach. It corresponds to the bare interaction for the dual particles. Analogously to the two-particle diagrammatics for the original electrons, one can formulate Bethe-Salpeter equations and parquet equation for the dual particles with $\Floc$ as the bare interaction. {Within the parquet approximation, this bare interaction is employed as the fully irreducible vertex of the dual fermions. From the fully irreducible vertex}  
 the full dual vertex $\Fd$ and the vertex functions $\Phid_r$ reducible in one of the three channels $r\in\{ph,\pht,pp\}$
\begin{equation}
\Fd^\kkq = \Floc^\vvw+\Phid_{ph}^{\kkq}+\Phid_{\pht}^{\kkq}+\Phid_{pp}^{\kkq}
	\label{eq:parquet}
\end{equation}
 are calculated. Here and in the following we use the tilde to denote quantities that are defined in terms of dual particles. 
Let us further define
the vertex functions $\Gammad_r$ that are irreducible with respect to a given channel $r$, and are therefore given by the complement $\Gammad_r=\Fd-\Phi_r$. 

{The Bethe-Salpeter equations give a relation between the reducible vertices $\Phid_r$ and the propagator of the dual fermions $\Gd$:} 
\begin{eqnarray}
\Phid_{ph}^{\kkq} &=&\sum_{k_1}\Fd^{kk_1q}\Gd_{k_1+q}\Gd_{k_1}\Gammad_{ph}^{k_1k'q} \; , \label{eq:bse1}\\
\Phid_{pp}^{\kkq} &=& -\frac{1}{2}\sum_{k_1}\Fd^{k (k_1+q) (k'-k_1)}\Gd_{k_1+q} \Gd_{k+k'-k_1}\nonumber\\
& &\times\;\;\Gammad_{pp}^{(k+k'-k_1) k' (q-k'+k_1)}. \label{eq:bse2}
\end{eqnarray}
Here and in the following, all four vector sums implicitly include, for brevity, a normalization factor $1/{(\beta N)}$, i.e., $\sum_k$ actually denotes  $1/{(\beta N)}\sum_k$ similar as in previous publications, e.g. Ref.~\onlinecite{RMPvertex}. Note that there is no need to introduce a separate equation for  $\Phid_{\pht}$ since it can be obtained from $\Phid_{ph}$ via the crossing symmetry \cite{rohringer2012}. 

In order to calculate a dual self-energy $\Sigmad$ out of the full {dual} vertex, the dual Schwinger-Dyson equation
\begin{eqnarray}
\Sigmad_k &=& -\sum_{k'}\Floc^{\vvw=0}\Gd_{k'}-\frac{1}{2}\sum_{k'q}\Floc^{\vvw} \nonumber\\
& & \times \Gd_{k'}\Gd_{k'+q}\Gd_{k+q}\Fd^{\kkq}
	\label{eq:schwinger}
\end{eqnarray}
is employed and for the propagator of the dual fermions $\Gd$ we can also formualte a Dyson equation
\begin{equation}
\Gd_k=\left[\Gd_{0,k}^{-1}-\Sigmad_k\right]^{-1}, 			\label{eq:dyson}
\end{equation}
where  $\Gd_{0,k}$ is the so-called non-interacting dual Green's function, which is the input to the dual fermion approach.
 

The non-interacting dual Green's function $\Gd_{0,k}$ is obtained as the difference between the $\mathbf k$-dependent and $\mathbf k$-averaged DMFT Green's function which can both be calculated from the DMFT self-energy $\Sigma_{\rm loc}$, cf.~Ref.~\onlinecite{RMPvertex}:
{}{
\begin{equation}
\Gd_{0,k}=\frac{1}{i\nu-\epsilon_{\veck}+\mu-\Sigma_{{\rm loc},\nu}} - \sum_{\bf k} \frac{1}{i\nu-\epsilon_{\veck}+\mu-\Sigma_{{\rm loc},\nu}} .
\end{equation}}
For the results presented, we keep $\Gd_{0,k}$ and  $\Floc$ fixed at their DMFT values, i.e., we do not do a so-called  ``outer'' self-consistency. As we will see below, the DF corrections to the self-energy are minute, justifying {\em a posteriori} that no   ``outer'' self-consistency is necessary.

The equations (\ref{eq:parquet}), (\ref{eq:bse1}), (\ref{eq:bse2}), (\ref{eq:schwinger}) and (\ref{eq:dyson}) can be employed in a self-consistent form (``inner'' self-consistency), where first the dual vertex $\Fd$ is built up from $\Floc$ and $\Gd_0$ to calculate $\Sigmad$ and the corresponding interacting $\Gd$, which are then used again in an updated calculation for $\Fd$. 

\subsubsection*{Postprocessing}
The resulting dual self-energy from the parquet equations is finally used as a nonlocal correction to the lattice Green's function $G$ of the real electrons:\footnote{Note, that in contrast to Ref.~\onlinecite{Rubtsov2008,Rubtsov2009} but in agreement with e.g. Ref~\onlinecite{ribic2016}, we use the DF self-energy directly as the physical self-energy without transformation formula between both self-energies. This is because the transformation formula only holds if diagrams from higher order vertices are included.\cite{Katanin2013} For the same reason, we directly use the full impurity vertex, without using the same transformation factors as for the self-energy to arrive the DF interaction, cf. Ref.~\onlinecite{Rubtsov2009}. In our calculations the DF self-energy is anyhow rather small so that these transformation formulas become just the identity up to very small corrections. }
\begin{equation}
G_k=\frac{1}{i\nu-\epsilon_{\veck}+\mu-\Sigma_{{\rm loc},\nu}-\Sigmad_k}.
\end{equation}

Using the results of the parquet DF formalism, also physical susceptibilities are calculated, namely the density-density correlation function or charge susceptibility $\chi_d$,
 \begin{eqnarray}
      \chi_{d,\vecq} 
       &=& \sum_{k}
          G_{q+k}G_{k}  \nonumber \\
       &&+ \sum_{k,k'}          
          G_{k'}G_{q+k}
          F^{kk'q}
          G_{q+k'}G_{k} \;,
  \end{eqnarray}
and the current-current correlation function $\chi_{jj}$,
 \begin{eqnarray}
      \chi_{jj,\vecq} 
       &=& 
          \left[\gamma_{\veck}^{\vecq}\right]^2
          G_{q+k}G_{k}  \nonumber \\
       &&+\sum_{k,k'}
          \gamma_{\veck}^{\vecq} \gamma_{{\veck}'}^{\vecq}
          G_{k'}G_{q+k}
          F^{kk'q}
          G_{q+k'}G_{k} \;.
  \end{eqnarray}
Here the full vertex function of the real fermions is approximated by the vertex function of the dual fermions, $F=\Fd$; $\gamma_{\veck}^{\vecq=0}=\partial\epsilon_{\veck}/\partial \veck$ denotes the dipole matrix elements given by the derivative of the energy-momentum relation in the Peierls approximation. From $\chi_{jj}$ at $\vecq=0$ the optical conductivity can be calculated,
 
\begin{equation} 
    \sigma(\omega) 
      = \mathrm{Re}\left \{ e^2 \lim_{\delta\rightarrow 0} \left[\frac{\chi_{jj,\vecq=0}(\omega+i\delta)
                  -\chi_{jj,\vecq=0}(i\delta)}{i(\omega+i\delta)} \right] \right\}.
\label{eq:ccCkw}
\end{equation}
Here, an analytic continuation to real frequencies is necessary, for which we employ the maximum entropy method described in the supplemental material of Ref.~\onlinecite{Kaufmann2018}.

\subsubsection*{Ladder DF}
\label{sec:model:ladder}
In addition to the full parquet DF calculation discussed above, we present results for a ladder DF approximation. This ladder approximation is employed in two different ways:

First, using the parquet formalism and code above, but restricting ourselves to the respective channel $r\in\{ph,pp\}$, i.e. Eq.~(\ref{eq:bse1}) or (\ref{eq:bse2}) and $\Fd=\Floc+\Phid_r$. Therefore, the ladder series is built up iteratively and a direct comparison with the parquet results is enabled in this way. In the case of the particle-hole ladder, both the $ph$- and the $\pht$-contributions are taken into account in the dual Schwinger-Dyson equation (\ref{eq:schwinger}) to recalculate the self-energy self-consistently.  For the $pp$-ladder series instead, $\Floc+\Phid_{pp}$ is employed in  the Schwinger-Dyson equation. This method is used to calculate the ladder results shown in Fig. \ref{fig:sigma_ladder_halffill}, \ref{fig:cond_parquet_U05_halffill}, \ref{fig:cond_parquet_U15_halffill}, \ref{fig:sigma_ladder_doped} and \ref{fig:cond_parquet_U05_doped}.

Second, using the exact expression for the dual vertex function in the ladder approximation, i.e. the geometric series. This second approach is employed in this paper only for the $pp$-ladder, as we want to isolate the effect of weak localization corresponding with such diagrams. The $pp$-ladder vertex function then reads

\vspace{-3mm}

\begin{equation}
\Fd^{\vvw}_{\vecq}=\frac{\Floc^{\vvw}}{1-\Floc^{\vvw}\widetilde{\chi}^0_q},
\label{eq:pp_ladder}
\end{equation}
where $\widetilde{\chi}_0$ is calculated in $pp$-notation from
\begin{equation}
\widetilde{\chi}^0_q=\sum_{k}\Gd_{q-k}\Gd_k.
\end{equation}
This second method is used to calculate the $pp$-ladder results in Fig. \ref{fig:cond_pp_halffill} and \ref{fig:cond_pp_doped} to confirm weak localization. {Diagrammatically, this describes the two directions a closed loop can be taken in by a particle returning to the same site. In disorder models, constructive interference between those paths leads to an increase of the amplitude for remaining at the same site and therefore a reduction of mobility.} As in the first implementation based on the parquet code, the self-energy is also recalculated self-consistently in this second approach.

\section{Results}
\label{sec:results}

We solve the DF parquet equations for the FKM on a $6\times 6$ square lattice with periodic boundary conditions using 20 (positive) Matsubara frequencies in the case of the parquet DF approach. {Note that the restriction to such small systems is a severe approximation, necessary due to the immense numerical effort of parquet calculations. A finite-size scaling analysis was not possible within the current numerical implementation. With other approaches, where either the frequency dependence is reduced~\cite{Astretsov2019} or a form-factor expansion is used~\cite{TUPS}, such analysis might become possible in the future.} 

In the case of the $pp$-ladder approximation, the reduced numerical effort allows us to study  a $32\times32$ square lattice and 40 (positive) Matsubara frequencies. The temperature at which most results are calculated is $T=0.06$. We present results both for the half-filled FKM at $n_c=n_f=0.5$, and for the conduction-electron doped  FKM with  occupations $n_c=0.15$ and $n_f=0.5$. Since we show quantities that depend on either real or Matsubara frequencies, we use in the following $\nu$ and $\omega$ to denote real frequencies and $\nu_n$ and $\omega_n$ for the Matsubara ones.

\subsection{Half-filled system}
\label{sec:halffill}

\subsubsection*{DMFT spectrum}
\begin{figure}[t]
 \includegraphics[width=0.85\linewidth]{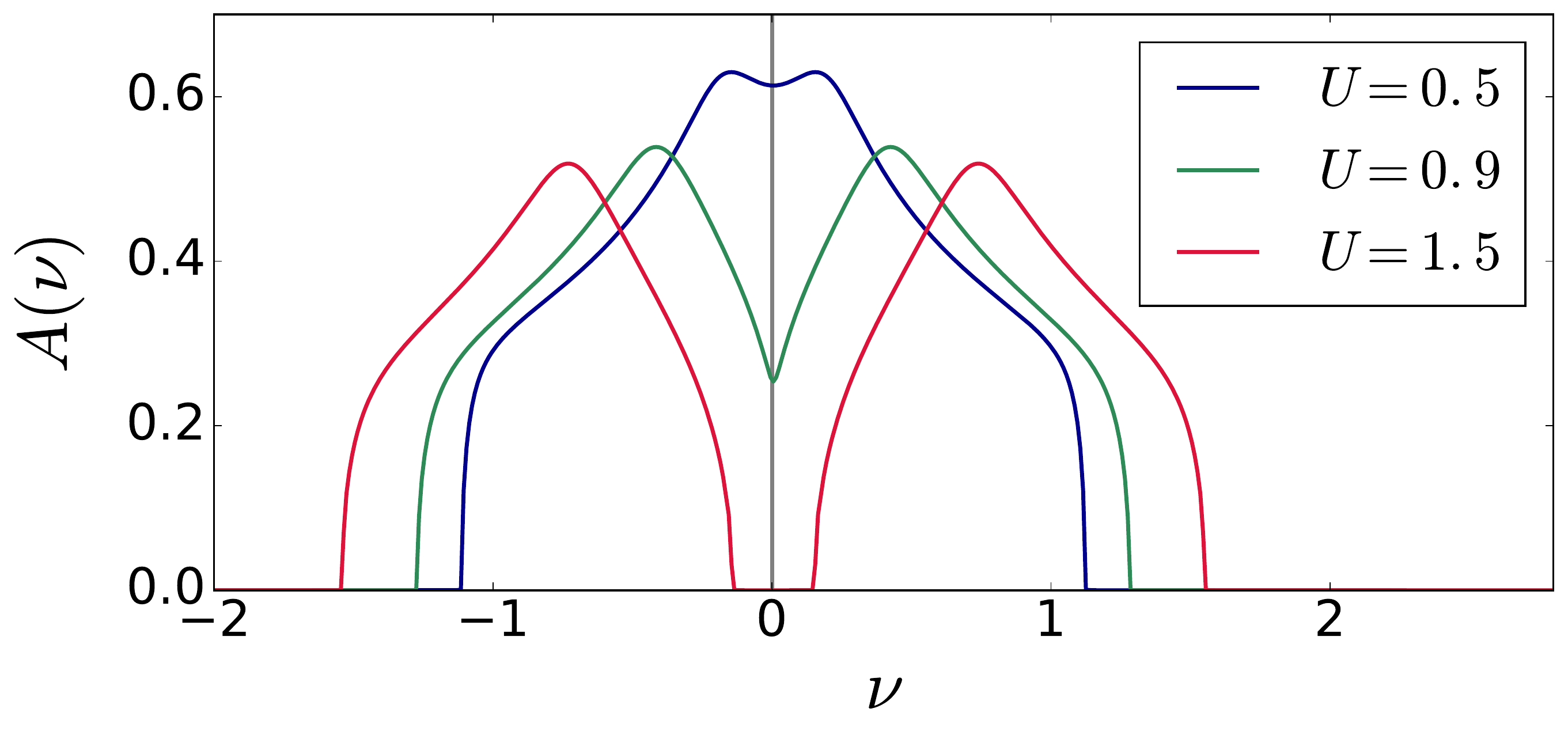}
   \caption{(Color online) DMFT spectral function $A(\omega)$ for the half-filled system, at $U=0.5$, $U=0.9$ and $U=1.5$. At $U=0.5$ and $U=0.9$, the system is still metallic. With increasing $U$, a gap forms and at $U=1$ the metal-to-insulator transition takes place; the spectrum is split into two subbands. This can be seen at $U=1.5$, where the system  is already insulating.}
   \label{fig:spectrum_halffill}
\end{figure}

In the case of the half-filled system, the chemical potential is fixed at $\mu=U/2$, and particle-hole symmetry holds. Electronic correlations are expected to have the maximum effect for this configuration and therefore it is in many cases most interesting to look at the system at half-filling especially when investigating the extension to nonlocal correlations. In DMFT for the two dimensional FKM at half-filling, a Mott-Hubbard-like metal-to-insulator transition occurs at $U=1$.\cite{Freericks2003} This can be seen in the DMFT spectral function $A(\nu)=-\frac{1}{\pi}\frac{1}{N}\sum_{\veck}\mathrm{Im}\ G(\veck,\nu)$ shown in Fig. \ref{fig:spectrum_halffill} on the real frequency axis for $U=0.5$, $U=0.9$ and $U=1.5$, where a gap is forming for increasing $U$ and at $U=1.5$ the spectrum is already split into two subbands. 
This DMFT solution, its vertex and bare dual Green's function serve as a starting point for the subsequent DF calculations. For these, we concentrate on one interaction ($U=0.5$) on the metallic side and one interaction ($U=1.5$) on the insulating side. 

\begin{figure*}
\begin{minipage}{0.7\linewidth}
 \includegraphics[width=\linewidth]{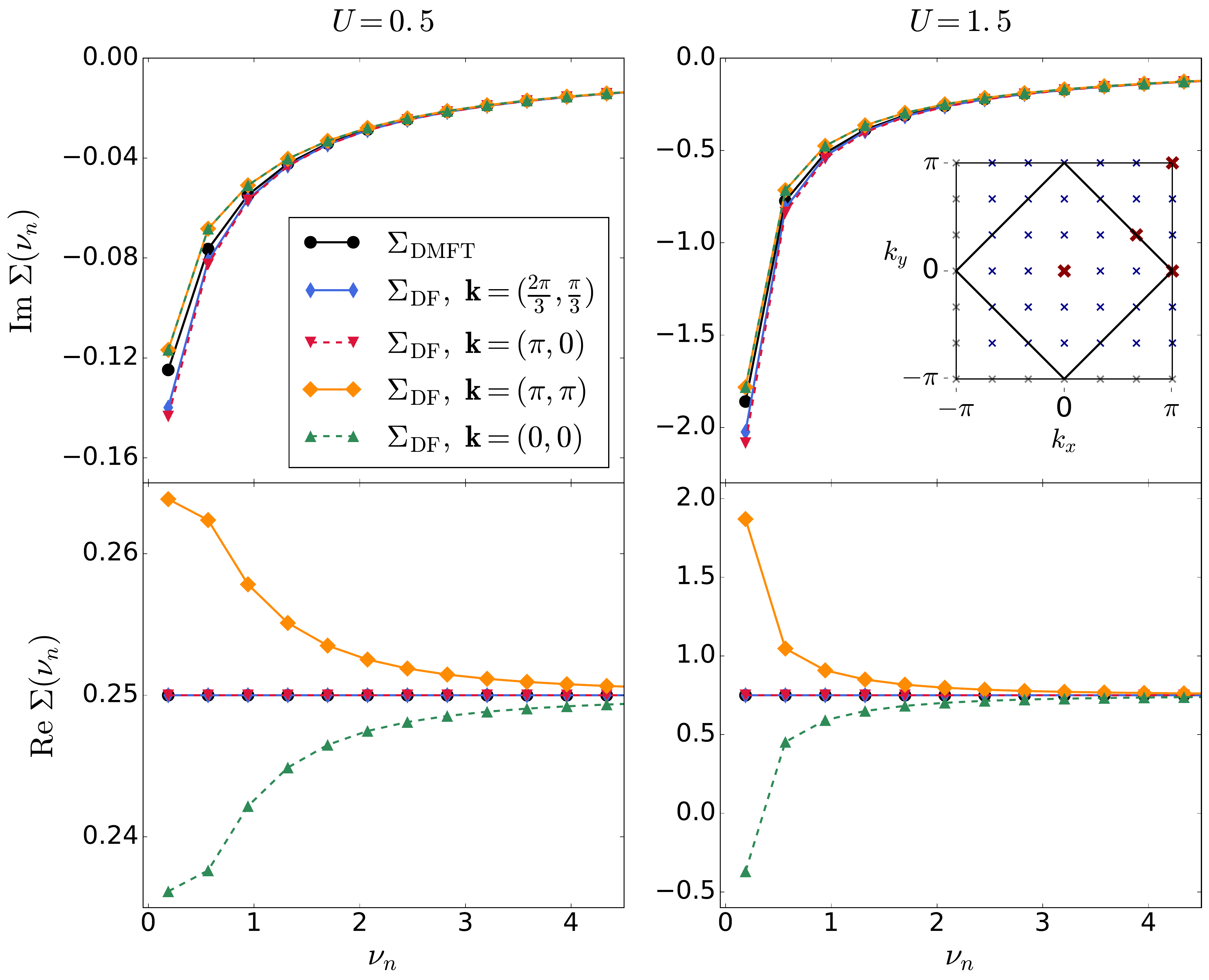}
\end{minipage}\hfill
\begin{minipage}{0.25\linewidth}\vspace{0pt}
   \caption{(Color online) Imaginary part (above) and real part (below) of the self-energy at $U=0.5$ (left) and $U=1.5$ (right)  at half-filling and $T=0.06$, as resulting from DMFT ($\Sigma_{\mathrm{DMFT}}$) and from the parquet DF approach ($\Sigma_{\mathrm{DF}}$). Right inset: Brillouin zone with the Fermi surface at half-filling (black line). The red crosses on the 6$\times$6 grid of $\bf{k}$-points denote the $\bf{k}$-points for which self-energies are shown in the main panel. \label{fig:sigma_parquet_halffill} }
\end{minipage}
  
\end{figure*}  

\subsubsection*{DF self-energy}
Fig.~\ref{fig:sigma_parquet_halffill} presents the results of the parquet DF self-energy in  comparison to the local DMFT self-energy at $T=0.06$ for the metallic and insulating system. Regarding the imaginary part of the self-energy, it can be seen that for the two $\veck$-points on the Fermi surface, $\veck=(\pi,0)$ and $\veck=(\frac{2\pi}{3},\frac{\pi}{3})$, the nonlocal corrections of the DF approach give a negative contribution to the DMFT self-energy, with $\SigmaDF$ being somewhat larger at $(\pi,0)$ (in absolute terms) than at $(\frac{2\pi}{3},\frac{\pi}{3})$. On the contrary, at $\veck=(\pi,\pi)$ and $\veck=(0,0)$ there is a positive contribution, reducing the absolute value of the self-energy. The real part of the self-energy is constant at $U/2$ in DMFT and in DF for $\veck$-points on the Fermi surface because of particle-hole symmetry; $\veck$-points outside the Fermi surface give positive, points inside negative DF corrections. That is the non-local DF self-energy pushes points further away in energy. The results look qualitatively similar both for the metal and the insulator, but note that the self-energy is an order of magnitude larger at $U=1.5$ compared to $U=0.5$.

\begin{figure}[t]
 \includegraphics[width=\linewidth]{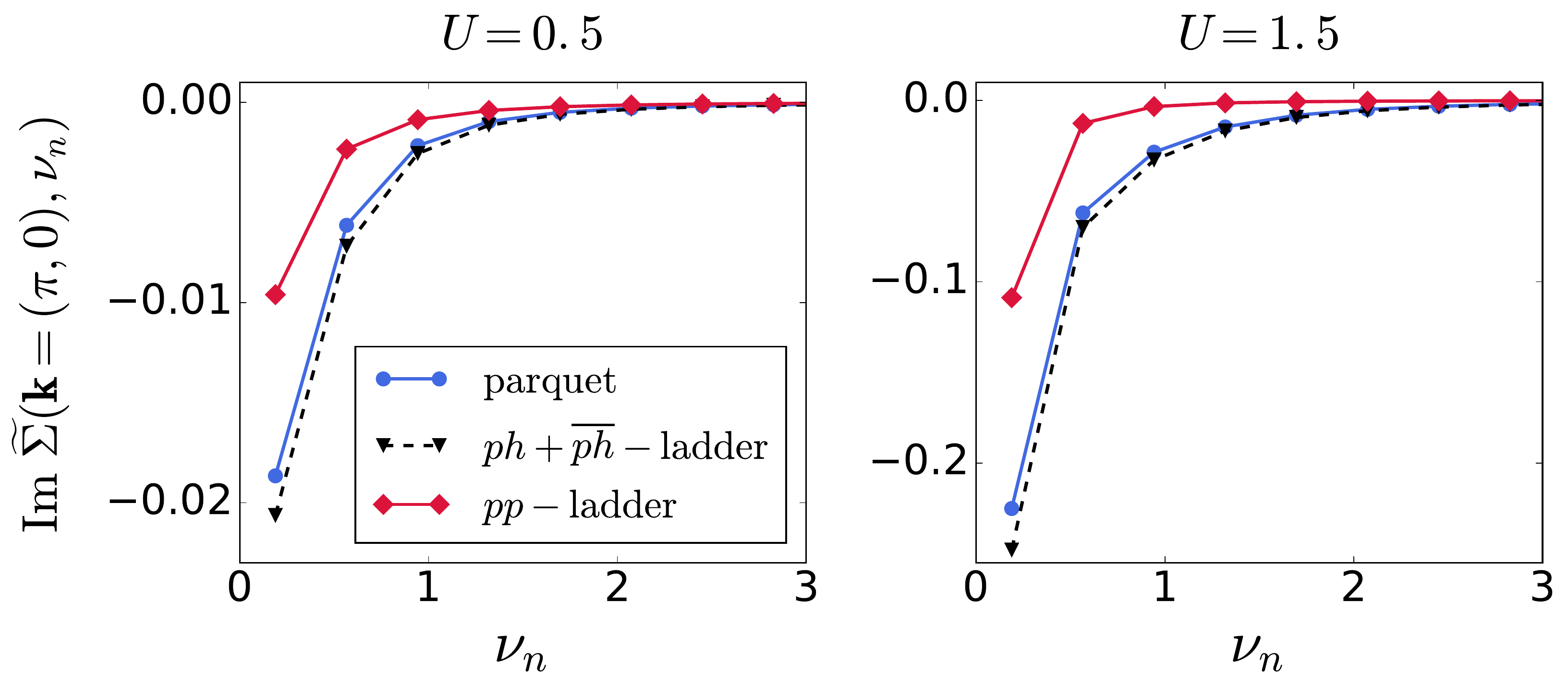}
    \caption{(Color online) Imaginary part of the dual self-energy at $U=0.5$ (left) and $U=1.5$ (right) at $\mathbf{k}=(\pi,0)$ as resulting from the full parquet DF approach (blue), a $ph+\pht$- (black dashed) and a $pp$-ladder (red) approximation for the half-filled system at $T=0.06$. The $ph$-ladder results containing also the $\pht$-contributions provide a good approximation to the dual self-energy as obtained from the full parquet calculation.}
    \label{fig:sigma_ladder_halffill}  
\end{figure}

In Fig. \ref{fig:sigma_ladder_halffill}  the parquet DF self-energy  is compared to a corresponding $ph+\pht$-ladder as well as to a $pp$-ladder DF approximation for the $(\pi,0)$-point. Both ladder series have been calculated iteratively, as described in section \ref{sec:model}. These results indicate that the physics of the FKM is dominated by $ph+\pht$-ladder diagrams, as the dual self-energy calculated in a simple ladder series approximates the results from the full parquet calculation very well. In contrast, $\Sigmad$ resulting from a $pp$-ladder calculation is considerably smaller.
Overall, the self-energy corrections are rather minute, at least for momenta on the Fermi surface. This justifies {\em a posteriori} that we do not need to recalculate the local vertex $\Floc$.

\begin{figure}[h]
 \includegraphics[width=\linewidth]{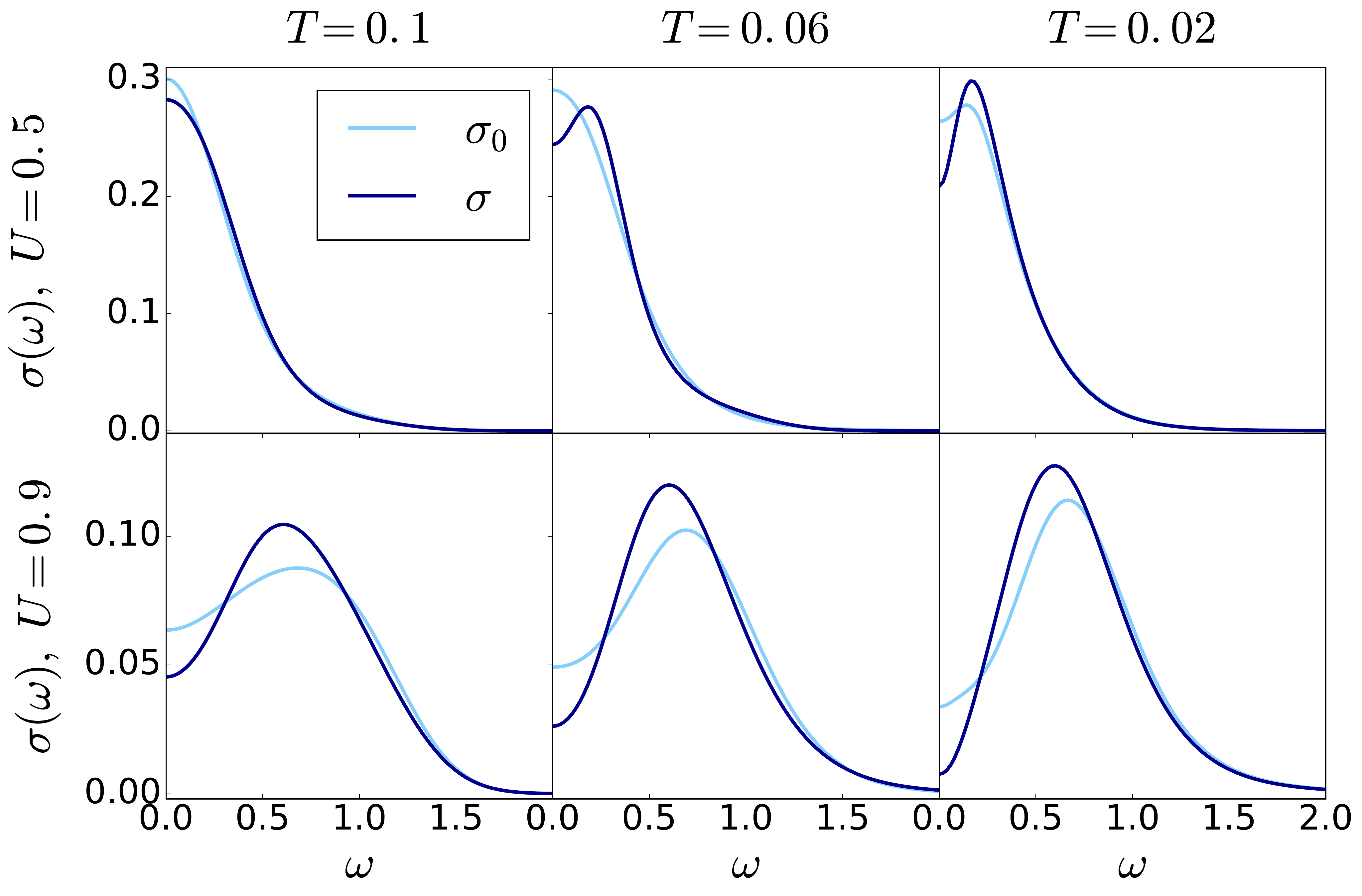}
   \caption{(Color online) Bubble term $\sigma_0$ and total optical conductivity $\sigma$ at $U=0.5$ (above) and $U=0.9$ (below)  as resulting from the $pp$-ladder approximation at $T=0.1$, $T=0.06$ and $T=0.02$ for a half-filled system. The effect of weak localization is clearly visible when employing only the $pp$-ladder.}
   \label{fig:cond_pp_halffill}
\vspace{-5mm}
\end{figure}

\subsubsection*{Optical conductivity}
\begin{figure}[h]
 \includegraphics[width=\linewidth]{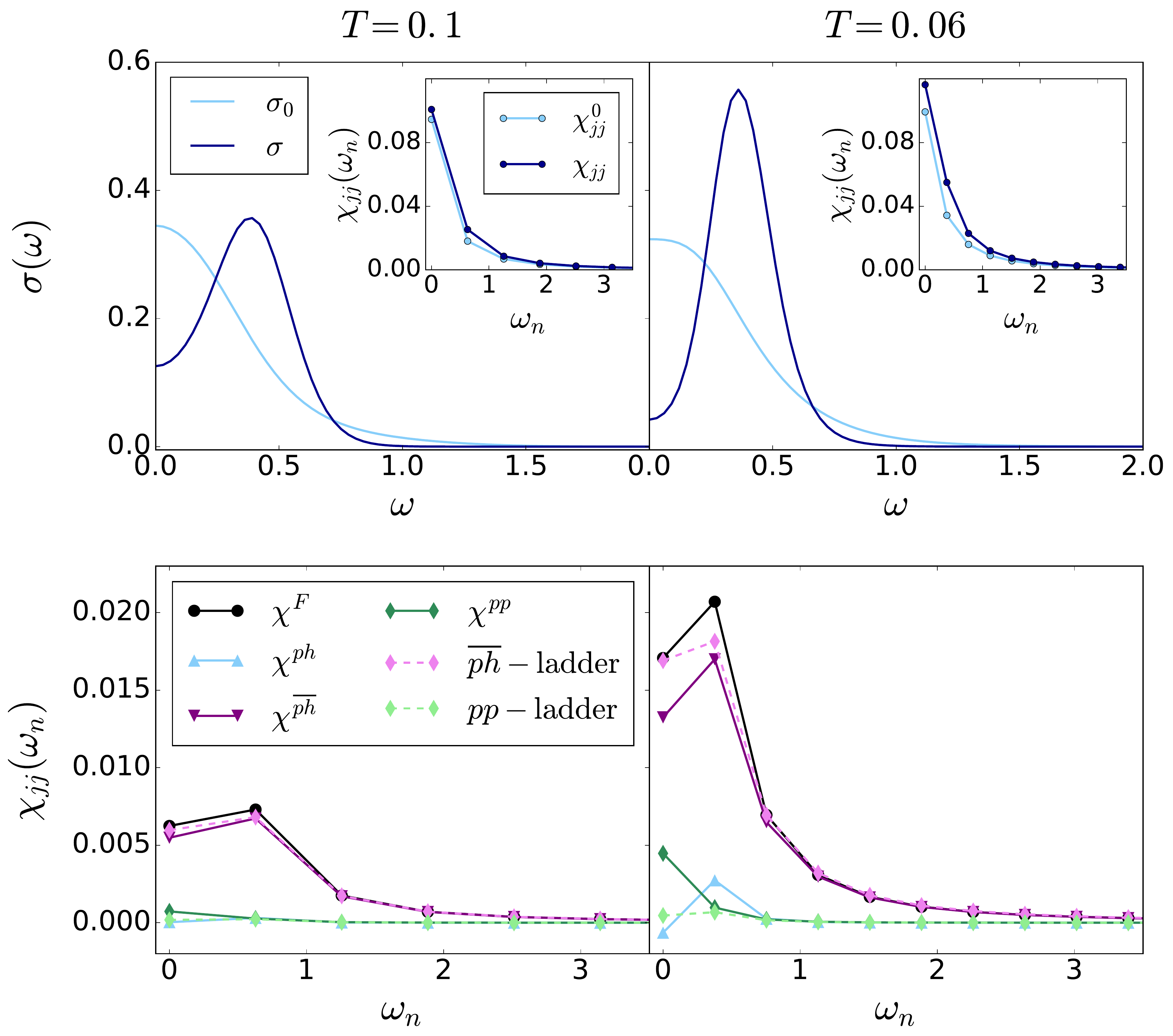}
    \caption{(Color online) Top: Optical conductivity for real frequencies (main panel) and the corresponding current-current correlation function in Matsubara frequencies (insets) for the half-filled FKM calculated now from parquet DF at $U=0.5$, $T=0.1$ and $T=0.06$. Shown are again the bare bubble ($\sigma_0$) and the full conductivity ($\sigma$) including vertex corrections. Insets: the respective current-current correlation function $\chi_{jj}^0$ and $\chi_{jj}$. Bottom: Corresponding vertex correction to the current-current correlation function $\chi_{jj}$ separated into $ph$, $\pht$ and $pp$ contributions. Additionally, the contribution of a $\pht$- and a $pp$-ladder are shown. As can be seen, the full parquet calculation shows even bigger effects compared to Fig. \ref{fig:cond_pp_halffill}. But these do not originate from the $pp$-channel, but the $\pht$-channel.}
    \label{fig:cond_parquet_U05_halffill}
\end{figure}
As the FKM is a model for annealed disorder, we may expect weak localization corrections to the conductivity, a physical phenomenon that is emerging at low temperatures. Weak  \cite{Altshuler1985} (and strong\cite{Abrahams1979}) {localization} emerge from Feynman diagrams in the $pp$-channel that cause a reduction of the optical conductivity at $\omega=0$, even though no gap is present in the one-particle spectrum. Such an effect and its increase with decreasing temperature can be seen indeed in Fig.  \ref{fig:cond_pp_halffill}, where we have restricted ourselves to these  $pp$-diagrams.\footnote{
 The results presented here are calculated in a $pp$-ladder approximation, using equation (\ref{eq:pp_ladder}) to calculate the full dual vertex and without updating the propagator of the real electrons to emphasize the diagrammatics of the $pp$-ladder without corrections in the propagator.}\footnote{For similar calculations of  vertex corrections in disordered systems, cf.~Refs.\onlinecite{Janis2010} and \onlinecite{Pokorny2013}.} Both the bare bubble term $\sigma_0$ and the total optical conductivity $\sigma$ are shown at real frequencies for two values of the interaction, $U=0.5$ and $U=0.9$, for which the one-particle DMFT and DF spectrum (which is essentially the same) is metallic. 
 At $U=0.5$ the bubble conductivity shows a typical Drude-like peak with maximum conductivity at $\omega=0$, however with a huge broadening because of the disorder scattering. In contrast at $U=0.9$, the system is close to the metal-to-insulator transition and therefore there is reduced optical weight at small frequencies in the bubble term due to the forming gap in the spectral function that can be seen in Fig. \ref{fig:spectrum_halffill}. For both interaction values, the vertex corrections from the $pp$-ladder yield a negative contribution to the conductivity for small frequencies, an effect that increases with decreasing temperature.
This is precisely the kind of physics expected from weak localization corrections. We can hence conclude that we are in a parameter regime where  conventional $pp$-diagrams yield weak localization corrections.

If we now employ the parquet equations  in Fig.~\ref{fig:cond_parquet_U05_halffill}  (top) instead of the mere $pp$-diagrams,  the behavior  is qualitatively very similar {as} for $U=0.5$. Quantitatively, the vertex corrections are however strongly enhanced: Now much of the optical spectral weight at $\omega=0$ is shifted towards higher frequencies, and a peak at around $\omega=0.4$ is forming. With lower temperature, the bubble conductivity itself is slightly reduced by the stronger nonlocal corrections to the Green's function; and the effect of the vertex corrections is further enhanced.

 However, the physical origin is a completely different one. This can be seen in Fig.~\ref{fig:cond_parquet_U05_halffill} (lower panel) where  we analyze from which channel ($ph$, $\pht$ and $pp$)
 the vertex corrections in the parquet equation emerge. That is, to obtain  Fig.~\ref{fig:cond_parquet_U05_halffill} (lower panel), the contributions of reducible vertices $\Phi_{ph}$, $\Phi_{\pht}$ or $\Phi_{pp}$  to the current-current correlation function have been calculated independently instead of the full vertex $F$. Inserting in Eq.~(\ref{eq:ccCkw}) instead of $F$ one of the summands: $\Phi_{ph/{\pht}/{pp}}$ we obtain the contributions from the respective channels:  $\chi^{ph}$, $\chi^{\overline{ph}}$, and $\chi^{pp}$.

Apparently the $\pht$-channel is the dominating one. Contributions from the $ph$- and $pp$-channel are rather small by contrast.  In addition to these parquet results, also results from a $\pht$- and $pp$-ladder are shown. Note that vertex corrections to the current-current correlation function in the $ph$-ladder vanish by symmetry. The difference of these simple ladder diagrams to diagrams emerging from the corresponding  $\Phi_r$  as calculated in parquet is the mixing of the channels, leading to a non-zero $\chi^{ph}$ and explaining the differences visible at $T=0.06$. 
Overall, we can conclude that the vertex corrections mainly stem from  $\pht$-contribution to the parquet equation, which in turn are essentially given by the $\pht$-ladder.

\begin{figure}[t]
 \includegraphics[width=\linewidth]{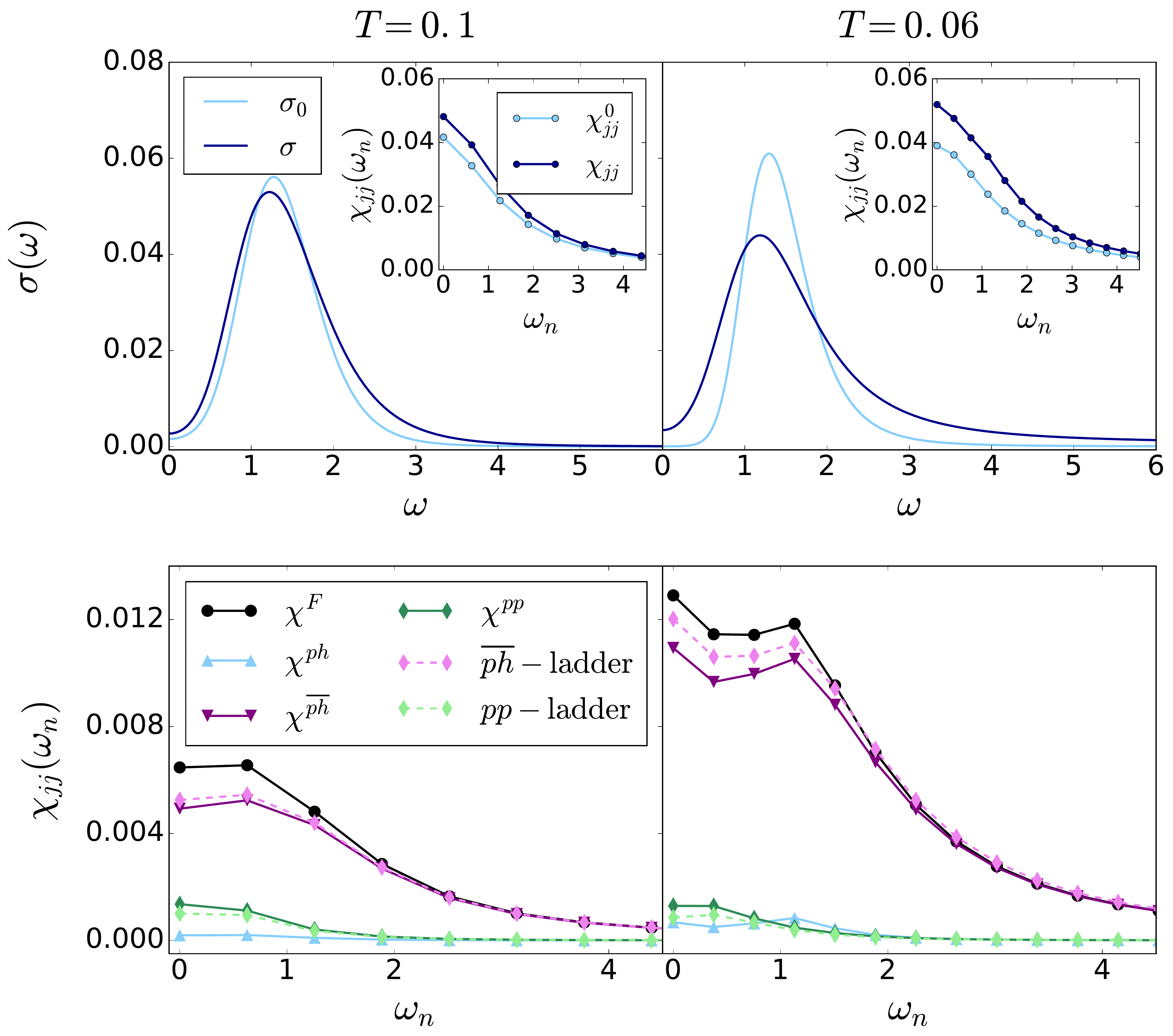}
    \caption{(Color online) Same as Fig. \ref{fig:cond_parquet_U05_halffill}, but for the insulating system at $U=1.5$.}
    \label{fig:cond_parquet_U15_halffill}
\end{figure}

 Fig.~\ref{fig:cond_parquet_U15_halffill} presents the same analysis but now for $U=1.5$. Here the bubble term $\sigma_0$ of the optical conductivity is centered around $\omega\approx U$, which corresponds to the distance of the peaks of the two subbands in the spectral function shown in Fig. \ref{fig:spectrum_halffill}. For low temperature, at $\omega=0$ the bubble conductivity vanishes because of the one-particle gap. It appears that in Fig.~\ref{fig:cond_parquet_U15_halffill} vertex corrections shift the optical weight  towards lower frequencies and that there is a finite weight at $\omega=0$ due to vertex corrections. But one has to keep in mind prospective uncertainties of the maximum entropy analytic continuation.
 Regarding the different contributions to the vertex corrections of $\chi_{jj}$, again $\pht$-reducible diagrams appearing in $\Phi_{\pht}$ are prevalent. 

\begin{figure}
 \includegraphics[width=0.8\linewidth]{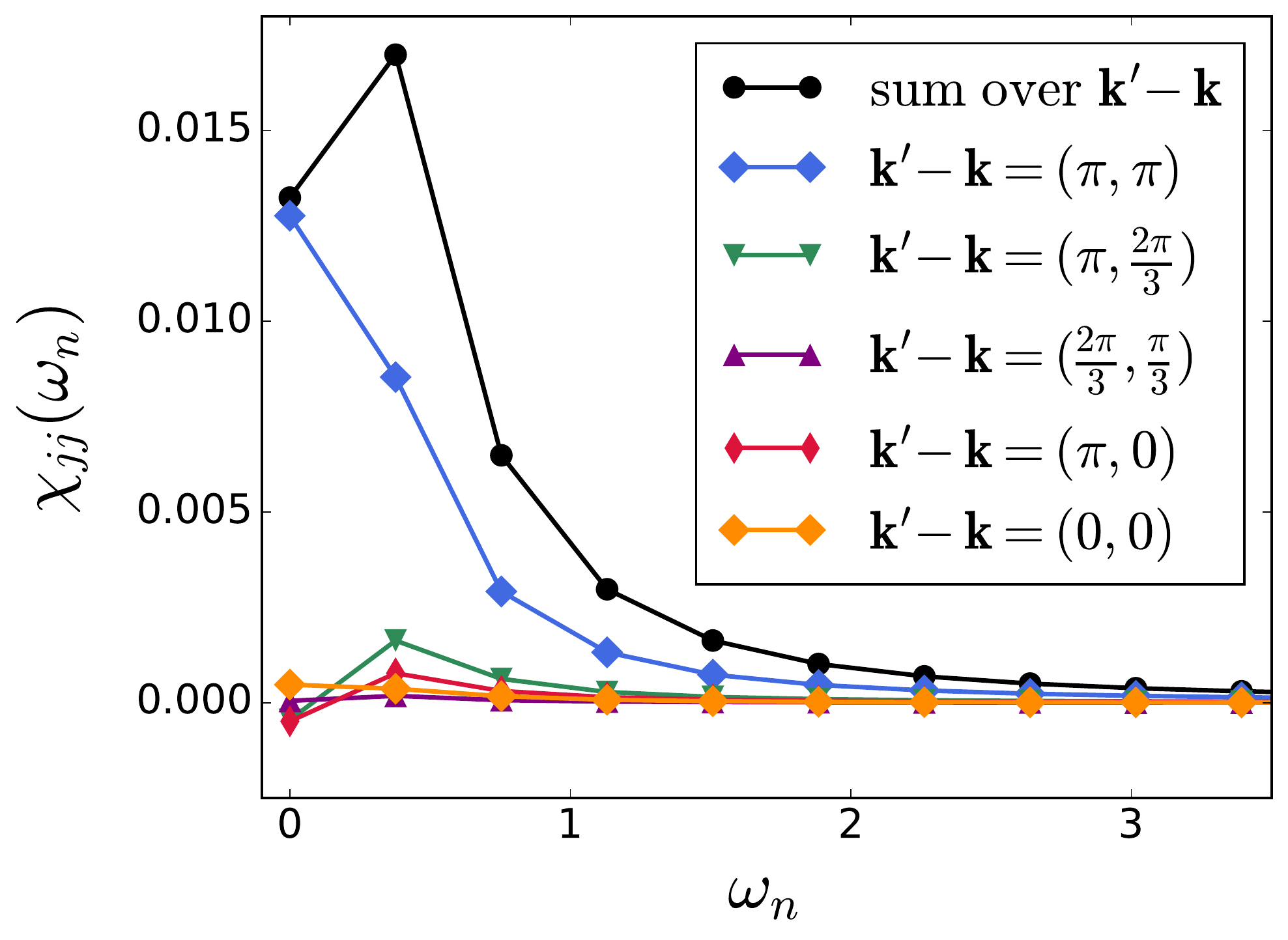}
    \caption{(Color online) Vertex corrections to the current-current correlation function $\chi_{jj}$ resulting from the $\pht$-channel (black) for representative momentum differences $\mathbf{k}'-\mathbf{k}$ for the half-filled FKM at $U=0.5$ and $T=0.06$.}
    \label{fig:chijj_pht_halffill}
\end{figure}

To better understand where the large vertex corrections from the $\pht$-channel come from, we analyze $\chi_{jj}$ as calculated from $\Phi_{\pht}$ further,  specifically  its contributions from different wave vectors $\veck'-\veck$. These contributions  for different momenta are shown in Fig. \ref{fig:chijj_pht_halffill}. Clearly the largest contribution stems from $\veck'-\veck=(\pi,\pi)$. At half-filling, this is the wave vector associated with CDW fluctuations, corresponding to a dominance of the  charge susceptibility $\chi_d(\vecq)$ at $(\pi,\pi)$.

\begin{figure}
 \includegraphics[width=\linewidth]{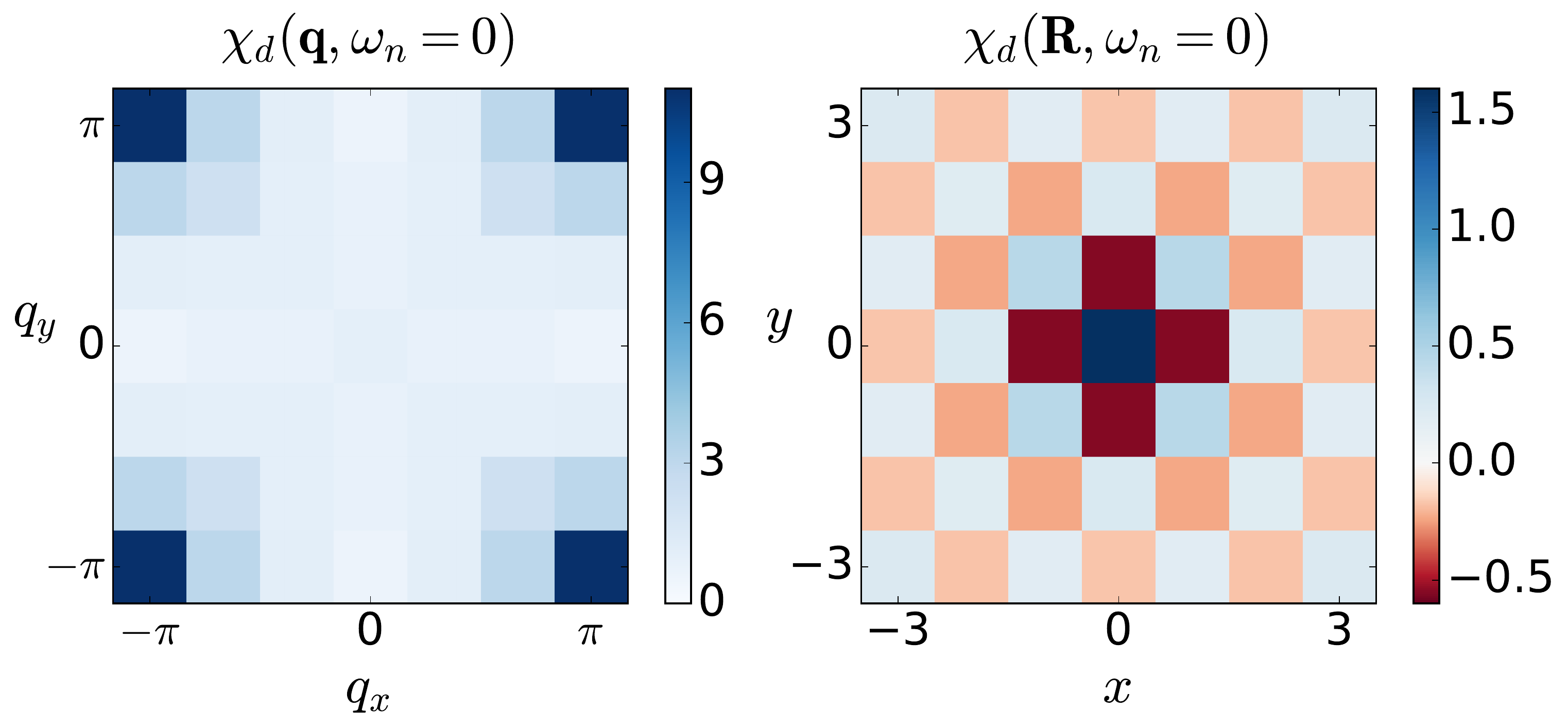}
    \caption{(Color online) Charge susceptibility $\chi_d$ at $\omega_n=0$ for a half-filled FKM at $U=0.5$ and $T=0.06$, both in momentum space (left) and real space (right). In momentum space, it can be seen that the dominating value lies at $\vecq=(\pi,\pi)$, which corresponds to the formation of the checkerboard structure which is also visible in real space.}
    \label{fig:chid_halffill}
\end{figure}

This is illustrated further in Fig. \ref{fig:chid_halffill}, which shows $\chi_d(\vecq)$  for the metallic system at $U=0.5$ and $T=0.06$  along with its Fourier transform, the charge susceptibility $\chi_d(\mathbf{R})$ in real space. A dominance of CDW fluctuations at  $(\pi,\pi)$ corresponds to a checkerboard structure in real space.

 All in all, we observe vertex corrections to the optical conductivity coming predominately from  $\veck'-\veck=(\pi,\pi)$ in the  $\pht$-channel. These contributions to optical conductivity can be interpreted as new polaritons, coined $\pi$-tons in Ref.~\onlinecite{kauch2019b}, where also first results for the FKM at $T=0.07$ (in units of $D\equiv 4t\equiv 1$) have been presented.

\subsection{$c$-doped system}
\label{sec:doped}

\subsubsection*{DMFT self-energy}

As CDW fluctuations are strongest for the  FKM  at half-filling,  one might expect that $\pht$-contributions are less relevant for the doped system and that the conventional weak localization picture with vertex corrections in the $pp$ channel reappears for the doped FKM.
Hence, 
we also present numerical results for the FKM on the square lattice  at an occupation of the mobile ($c$) electrons  $n_c=0.15$,
and of  localized $f$ electrons  $n_f=0.5$. To this end, we fix the chemical potential $\mu$ to the value  at which the DMFT solution yields the doping $n_c=0.15$. That is, $\mu$ is held constant throughout the parquet DF calculation, and therefore the occupation as resulting from the DF approach is changed compared to the corresponding DMFT solution. However, as we will see below, this change in occupation is minute.

In Fig. \ref{fig:spectrum_doped} we show again
the DMFT spectral function $A(\nu)$ for the $c$-doped system at $U=0.5$, $U=0.9$ and $U=1.5$. The behavior of the spectrum splitting into two subbands with increasing interaction strength is the same as at half-filling, but at this low occupation, the Fermi level always lies within the lower subband and therefore the system retains its metallic character at these parameters.

\begin{figure}
 \includegraphics[width=\linewidth]{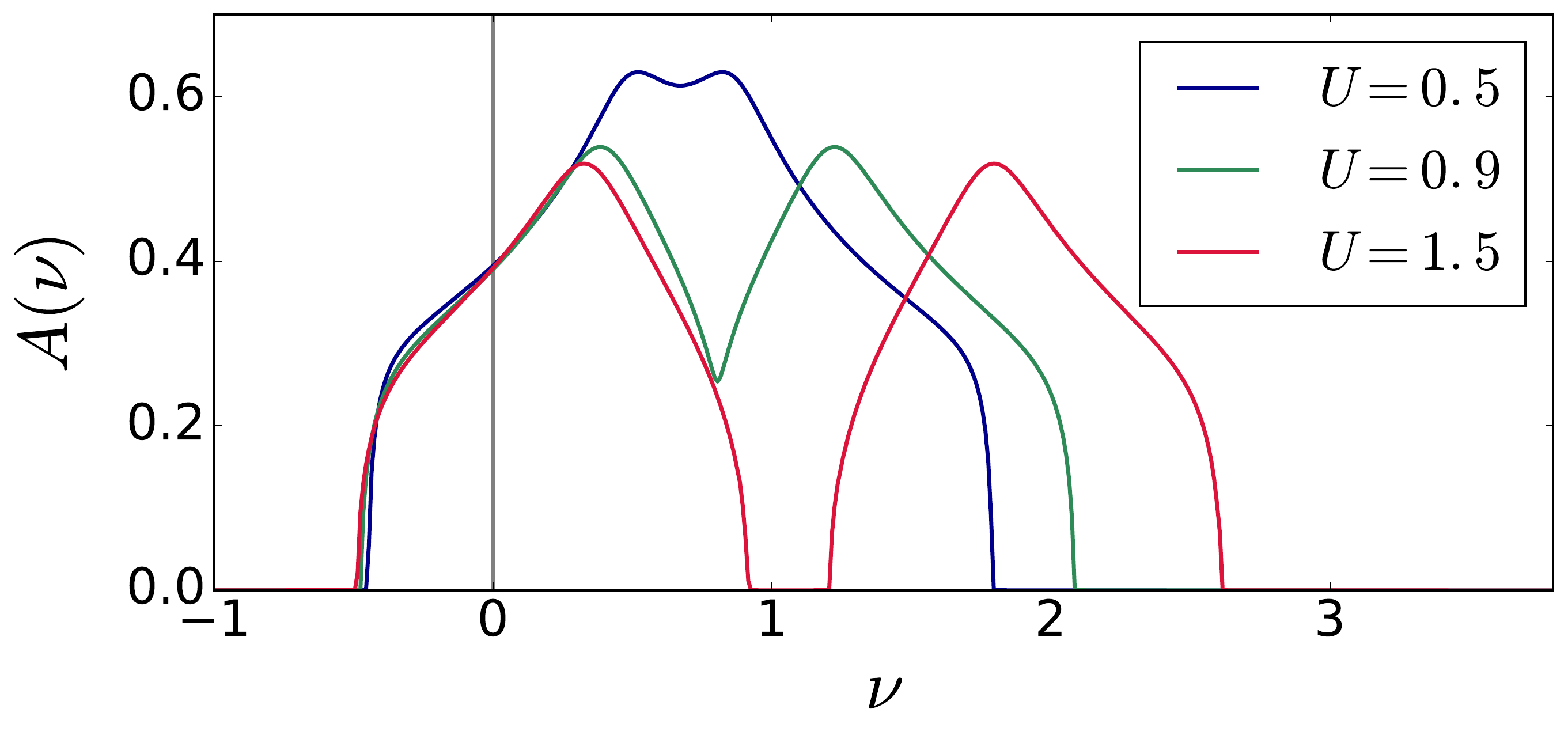}
   \caption{(Color online) DMFT spectral function $A(\nu)$ of the FKM at filling $n_c=0.15$ and $n_f=0.5$. For all three $U$ shown, the system remains metallic, even when the spectrum is split into two subbands at $U\geq 1$.}
   \label{fig:spectrum_doped}
\end{figure}

\subsubsection*{DF self-energy}
\begin{figure}
 \includegraphics[width=\linewidth]{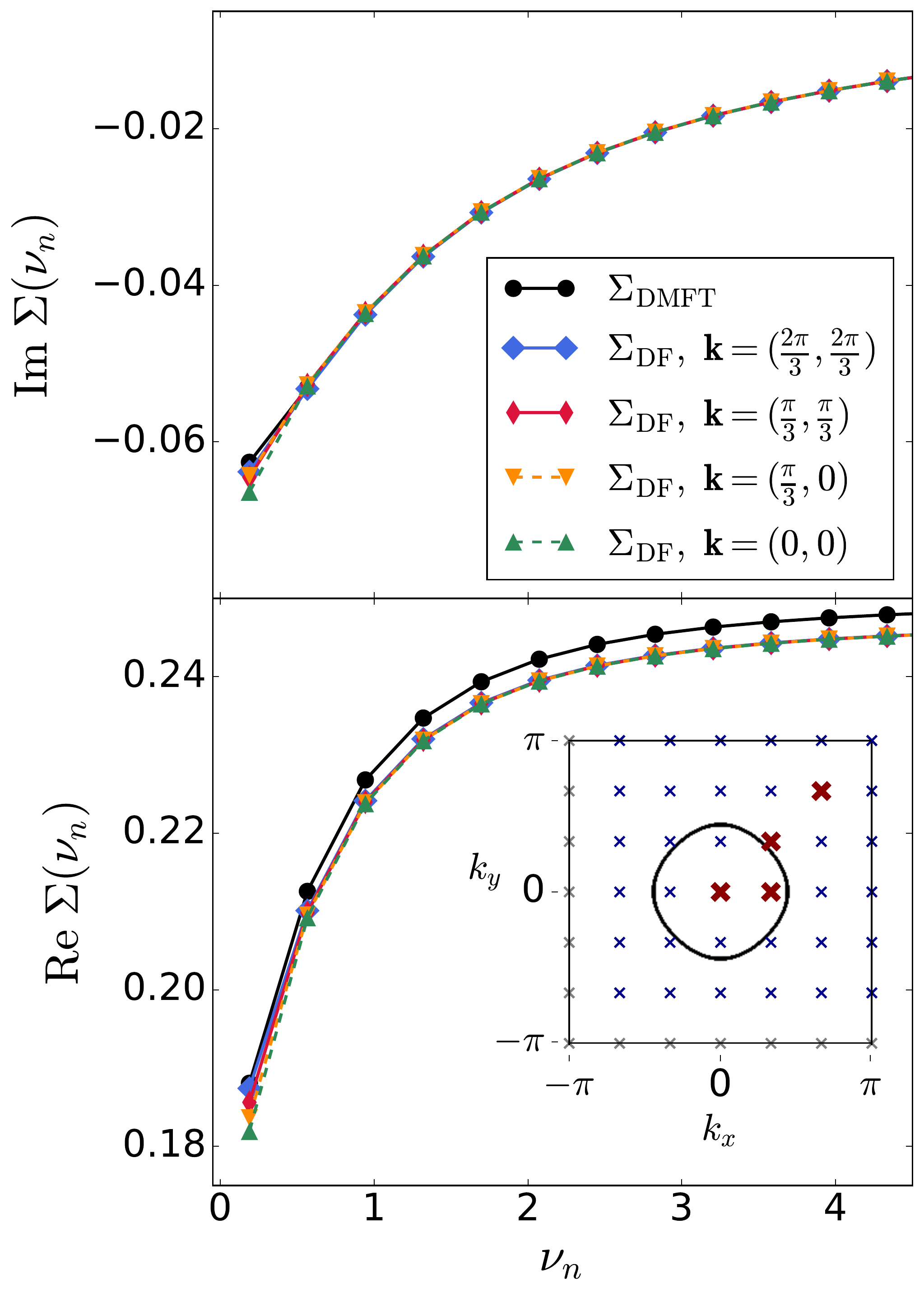}
   \caption{(Color online) Imaginary part (left) and real part (right) of the self-energy at $U=0.5$ and $T=0.06$  obtained in DMFT ($\Sigma_{\mathrm{DMFT}}$)  and parquet DF ($\Sigma_{\mathrm{DF}}$) for the FKM at filling $n_c=0.15$, $n_f=0.5$. The nonlocal corrections resulting from the parquet DF approach are smaller at this filling compared to the half-filled case in Fig. \ref{fig:sigma_parquet_halffill}.
   Right inset: Brillouin zone with the Fermi surface in DMFT for the $c$-doped system (black line).}
   \label{fig:sigma_parquet_doped}
\end{figure}

The results of the parquet DF approach at $c$-doping are compared to the DMFT calculation for the self-energy in Fig. \ref{fig:sigma_parquet_doped} for three different momenta $\veck$. The $(0,0)$-point lies of course well inside the Fermi surface at the filling of $n_c=0.15$, $\veck=(\frac{\pi}{3},\frac{\pi}{3})$ lies very close to it and $\veck=(\frac{2\pi}{3},\frac{2\pi}{3})$ lies outside of the Fermi surface.  The nonlocal corrections are much smaller than they are at half-filling (Fig. \ref{fig:sigma_parquet_halffill}), which can be expected as nonlocal correlations show the largest effect at half-filling. The resulting occupation in the DF calculation is  $n_c=0.1542$, only slightly different from the DMFT value $n_c=0.15$.\footnote{The change in the occupancy of the $c$ electrons is also reflected in the offset in the real part of the DF self-energy in Fig. \ref{fig:sigma_parquet_doped} that differs from the offset in DMFT.} 

\begin{figure}
 \includegraphics[width=0.7\linewidth]{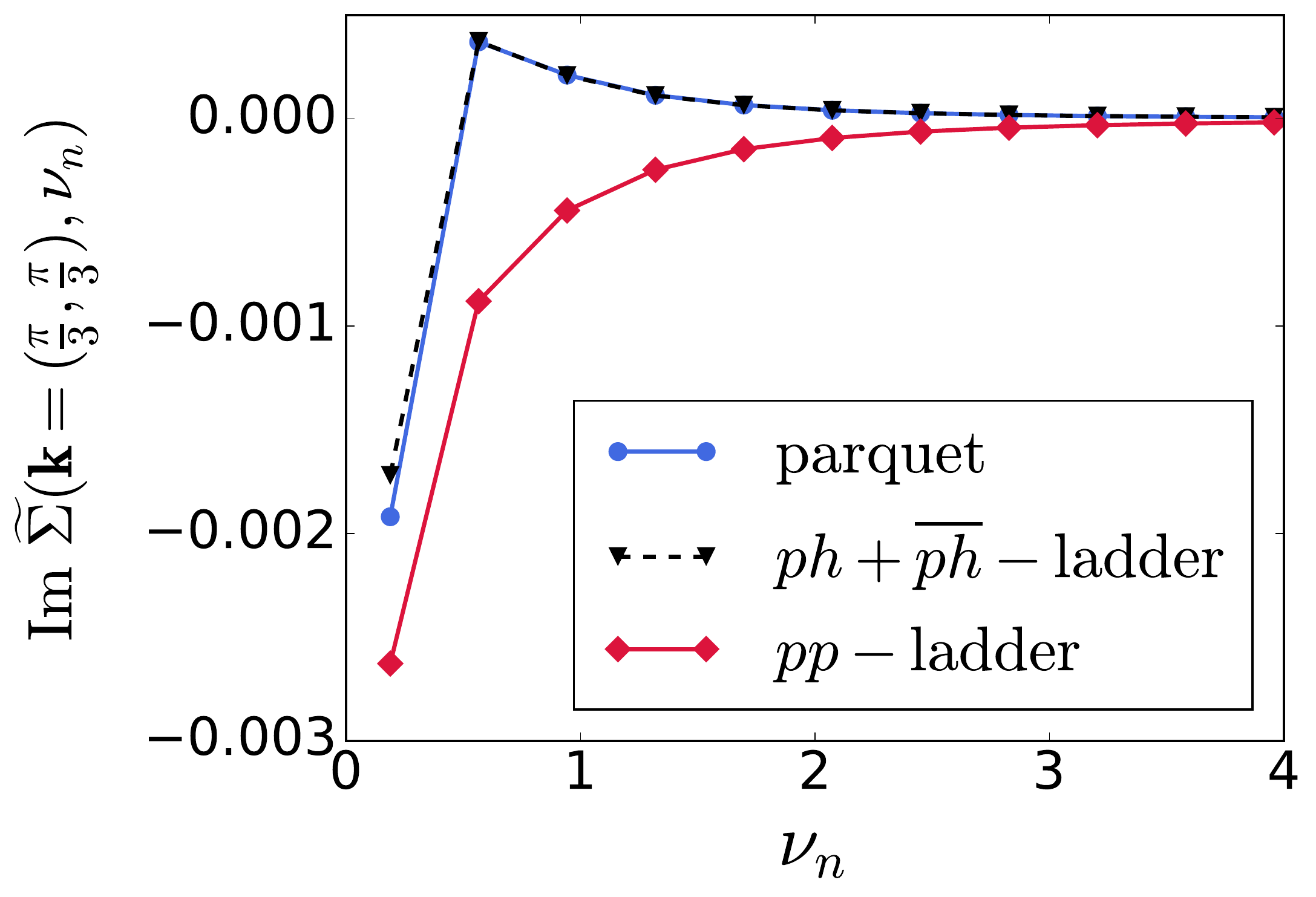}
    \caption{(Color online) Imaginary part of the dual self-energy at $U=0.5$, $T=0.06$ at $\mathbf{k}=(\frac{\pi}{3},\frac{\pi}{3})$ as resulting from the full parquet DF approach, a $ph+\pht$-  and $pp$-ladder approximation for an occupation $n_c=0.15$.  As in the half-filled case (Fig. \ref{fig:sigma_ladder_halffill}), the $ph$-ladder approximates the full parquet results very well, indicating the dominance of the $ph$-channel also in the doped FKM.} 
    \label{fig:sigma_ladder_doped} 
\end{figure}

As in the half-filled system, also in the $c$-doped system the $ph$-channel remains the dominating one. This can be seen in Fig. \ref{fig:sigma_ladder_doped}, where the dual self-energy from the full parquet calculation is compared to corresponding ladder approximations at $\mathbf{k}=(\frac{\pi}{3},\frac{\pi}{3})$. Again, the self-energy containing the $ph$- and the $\pht$-ladder approximates the parquet results very well, whereas the self-energy in the $pp$-ladder shows a qualitatively different behavior. 

\subsubsection*{Optical conductivity}

\begin{figure}
 \includegraphics[width=\linewidth]{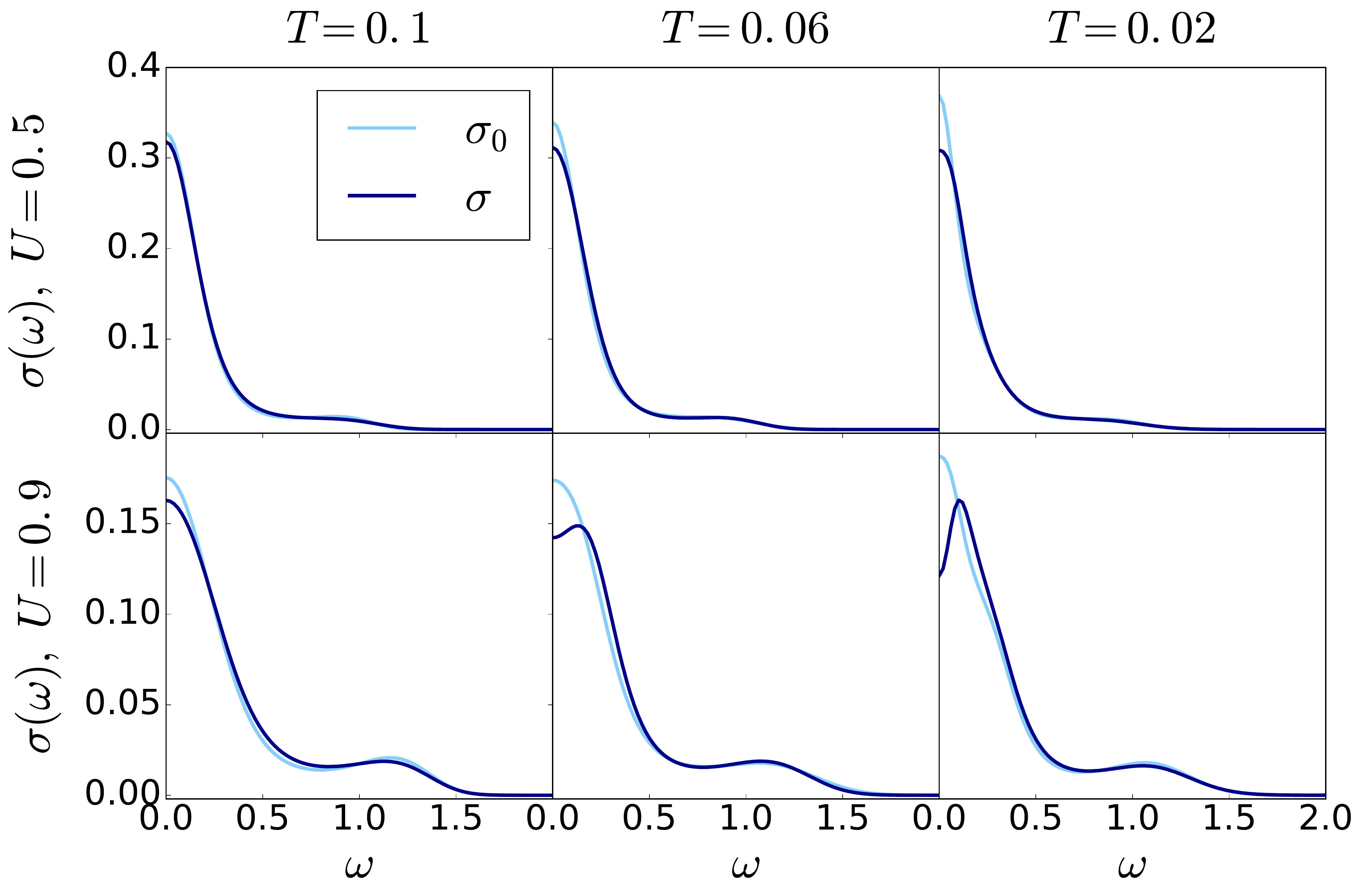}
   \caption{(Color online) Bubble term $\sigma_0$ and total optical conductivity $\sigma$ at $U=0.5$ (above) and $U=0.9$ (below)  as resulting from the $pp$-ladder approximation at $T=0.1$, $T=0.06$ and $T=0.02$ for the $c$-doped FKM, $n_c=0.15$. The effect of weak localization is clearly visible when employing only the $pp$-ladder.}
   \label{fig:cond_pp_doped}
\end{figure}

As in the case of half-filling we again first study, in Fig. \ref{fig:cond_pp_doped}, the vertex corrections to the optical conductivity  for the doped FKM as obtained from only  the $pp$-ladder, which can be associated with weak localization corrections.  
At the $c$-electron occupation $n_c=0.15$, the Drude-like peak in the bubble term $\sigma_0$ is accompanied by a small side peak corresponding to transitions from the Fermi level to the upper subband in the spectral function. Including the $pp$-ladder vertex corrections, optical spectral weight is reduced at and around $\omega=0$, as to be expected from weak localization corrections.

\begin{figure}[h]
 \includegraphics[width=\linewidth]{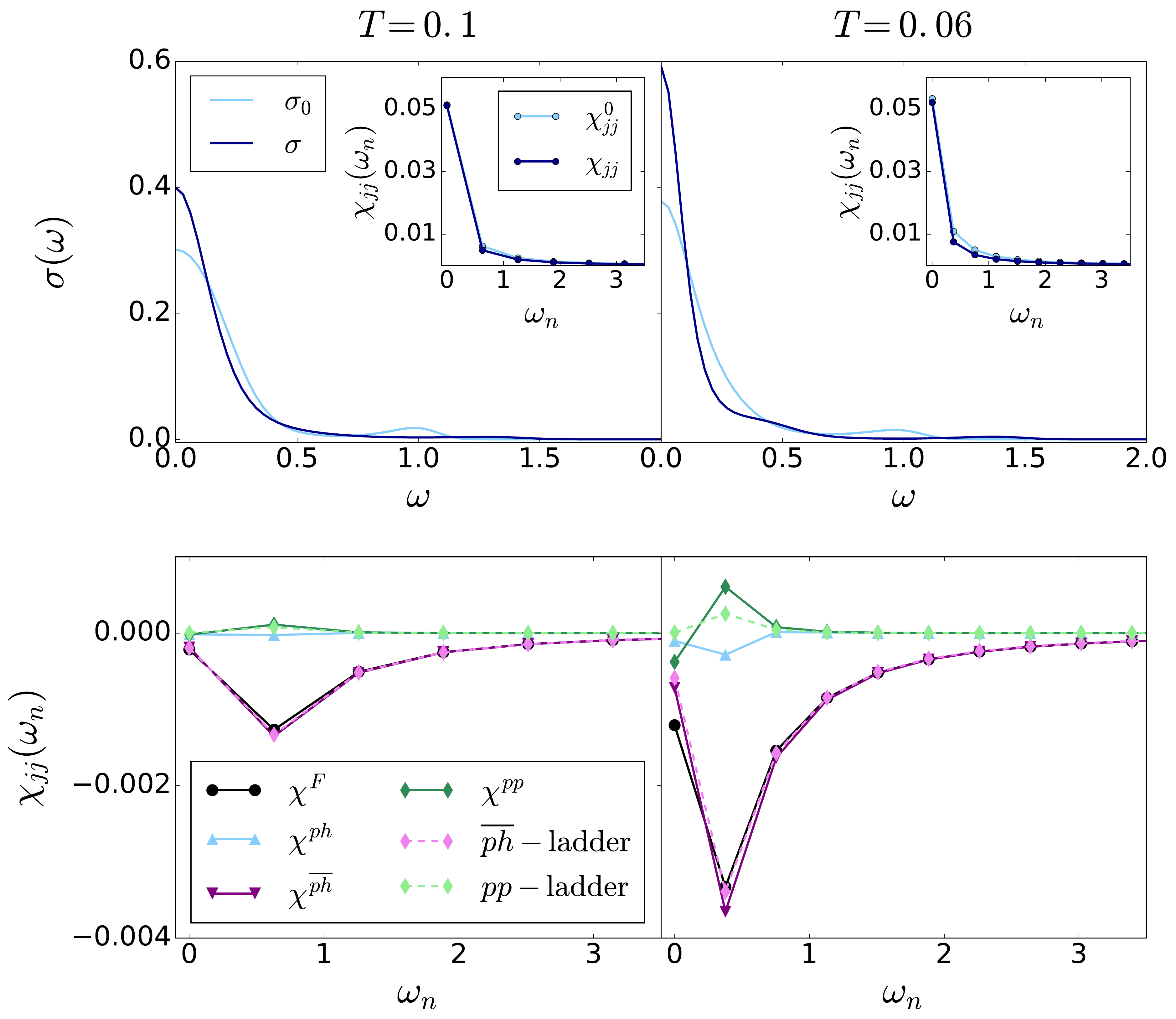}
    \caption{(Color online) Top: Optical conductivity for real frequency (main panel) and the corresponding current-current correlation function in Matsubara frequencies (insets) for the filling $n_c=0.15$ at $U=0.5$ and at $T=0.1$ and $T=0.06$, showing the bare bubble ($\sigma_0$) and the full conductivity ($\sigma$) including vertex corrections (in the insets $\chi_{jj}^0$ and $\chi_{jj}$ respectively). Bottom: Corresponding vertex correction to the current-current correlation function $\chi_{jj}$ separated into $ph$, $\pht$ and $pp$ contributions. Also the contribution of a $\pht$- and a $pp$-ladder are shown. As can be seen, the full parquet calculation show very different effects compared to Fig. \ref{fig:cond_pp_doped}.} 
    \label{fig:cond_parquet_U05_doped}
\end{figure}

The optical conductivity in the full parquet DF approach is shown in Fig. \ref{fig:cond_parquet_U05_doped} at $U=0.5$. As opposed to the results of the $pp$-ladder approximation and the results in the half-filled case (Fig. \ref{fig:cond_parquet_U05_halffill}), the vertex corrections in the full parquet calculation now lead to an increase of the optical conductivity at low frequencies, the small side peak is suppressed. The corresponding vertex contribution to the current-current correlation function shown in the lower panel of Fig. \ref{fig:cond_parquet_U05_doped} is negative and therefore has opposite sign to the bubble shown in the inset in the upper panel. 

The largest contribution stems, as for the half-filled system,  again from the $\pht$-channel. These are also at the origin of the negative current-current correlation function, whereas the contribution from the $pp$-channel in parquet DF and the $pp$-ladder yield positive vertex corrections to the current-current correlation function.

\begin{figure}[b]
 \includegraphics[width=0.7\linewidth]{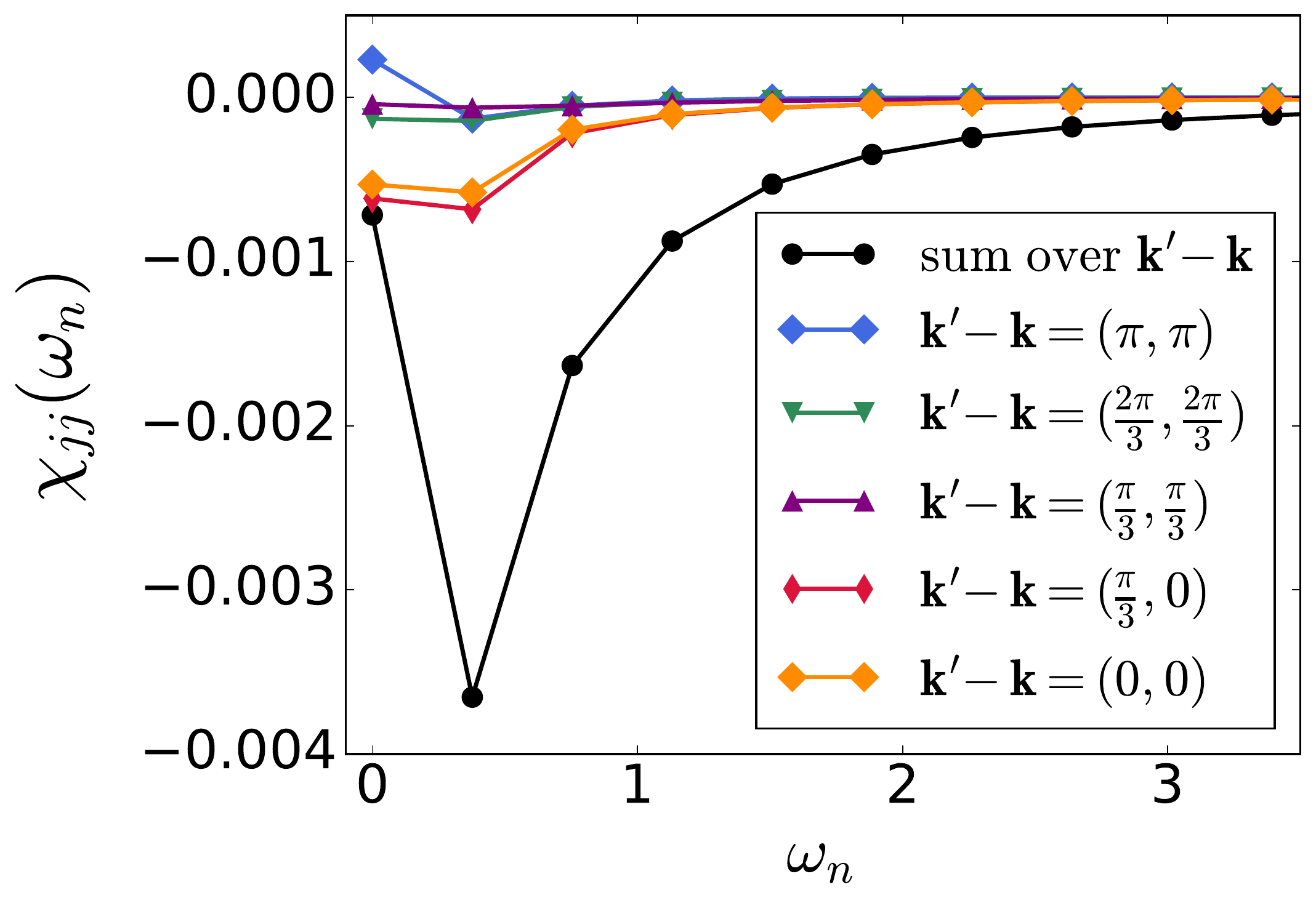}
    \caption{(Color online) Vertex corrections to the current-current correlation function $\chi_{jj}$ resulting from the $\pht$-channel in the parquet DF calculation (black) for different representative $\mathbf{k}'-\mathbf{k}$ for the doped FKM at doping $n_c=0.15$, $n_f=0.5$, interaction $U=0.5$ and $T=0.06$.}
    \label{fig:chijj_pht_doped}
\end{figure}

When we investigate in Fig. \ref{fig:chijj_pht_doped} the $\veck'-\veck$-dependence of the vertex part of the current-current correlation function stemming from the $\pht$-channel, we determine that it is the $(\frac{\pi}{3},0)$-point that yields the largest contribution at filling $n_c=0.15$. However, for the doped FKM other contributions are only slightly smaller, so that the current-current correlation obtained with the reducible vertex $\Phi_{\pht}$ shows not such a distinct connection to a single momentum as for the half-filled system (Fig. \ref{fig:chijj_pht_halffill}). 

\begin{figure}
 \includegraphics[width=\linewidth]{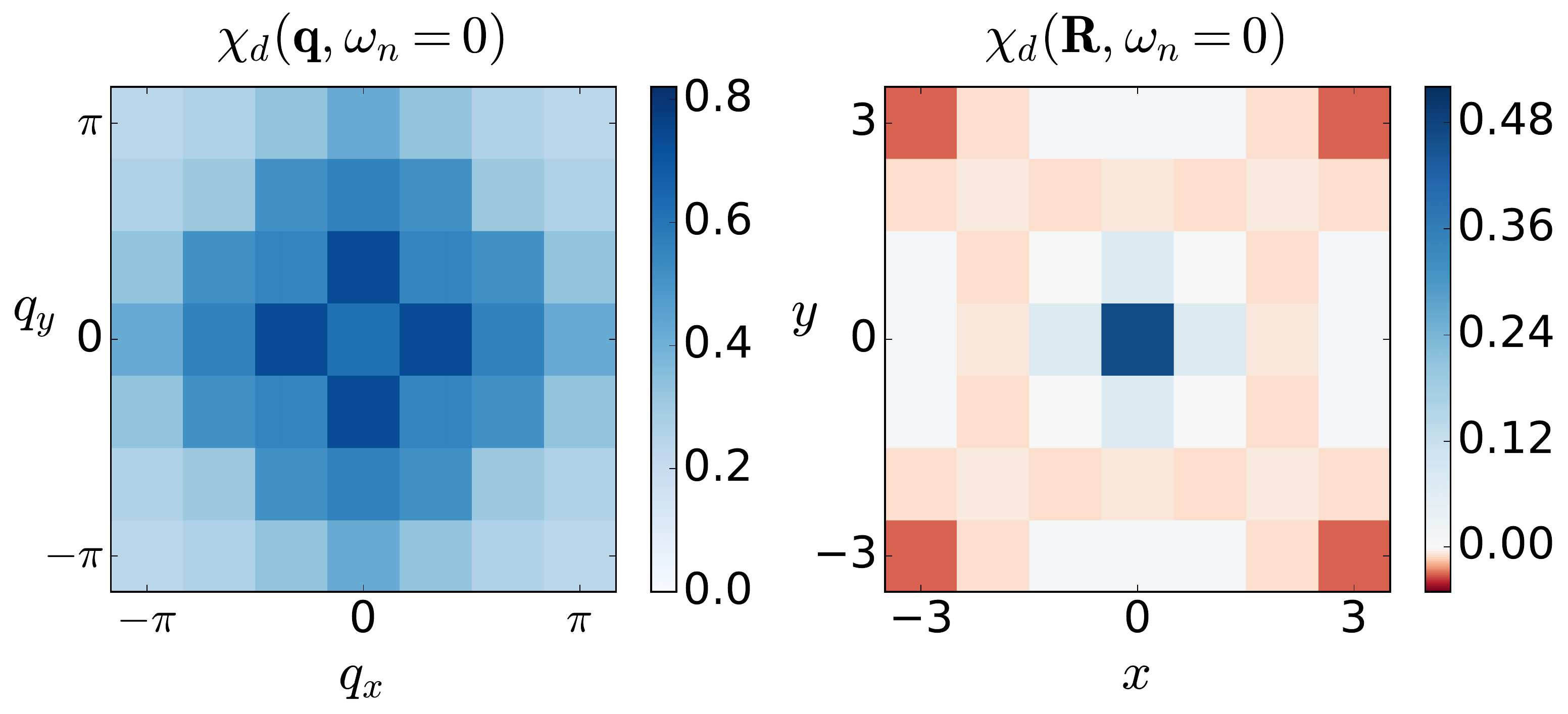}
    \caption{(Color online) Charge susceptibility $\chi_d$ at $\omega_n=0$ for the $c$-doped FKM at $n_c=0.15$, $U=0.5$, $T=0.06$, both in momentum space (left) and real space (right). The visible features with $\vecq=(\frac{\pi}{3},0)$ as strongest point are not as distinctive as in the half-filled system (compare Fig. \ref{fig:chid_halffill}). In real space, $\chi_d$ consists mostly of the on-site correlation of a $c$-electron, as the amplitude decreases rapidly around the origin.}
    \label{fig:chid_doped}
\end{figure}

The vector $\veck'-\veck=(\frac{\pi}{3},0)$ is also for the doped FKM the momentum where the static charge susceptibility $\chi_d$  is strongest in the doped system. This is visible in Fig. \ref{fig:chid_doped}, where $\chi_d$ for the whole lattice in momentum and real space is shown at $T=0.06$ and $U=0.5$.

\section{Conclusion}
\label{sec:conclusion}

In this paper, effects of nonlocal correlations in addition to the local ones as described in DMFT have been analyzed. To this end, a full parquet dual fermion approach was employed to obtain such nonlocal vertex corrections for the Falicov-Kimball model. This goes beyond previous investigations using ladder approaches \cite{ribic2016,antipov2014,yang2014}. Our code can be found at GitHub~\cite{code_fkm} with implementation details given in Ref.~\onlinecite{Astleithner_Thesis}

As expected, $ph + \pht$-diagrams corresponding to charge density wave fluctuations are dominating in the FKM, and      such a ladder approach already is a good approximation for the numerically very cumbersome parquet approach.
However, the  $ph$-channel cannot directly couple to light, since the former has a momentum $(\pi,\pi)$ at half filling and $(\pi/3,0)$ for the doped case considered in this paper, whereas light transfers momentum ${\mathbf q}=0$ to the solid.
 One may expect  weak localization corrections in the $pp$-channel instead to play an important role for vertex corrections to the conductivity. In this situation, {a method unbiased in the choice of diagrams employed}, such as the parquet approach, which takes all of the different fluctuations into account is necessary.

{}{We find that the   $pp$-contributions, aka weak-localization corrections, are present if we consider the  $pp$-ladder only. However, if we include all scattering channels other effects, originating from the $\pht$-channel, dominate. This $\pht$-channel can couple light to charge density wave fluctuations by exciting  two electrons and two holes whose momentum differences matches the dominating momentum of the charge density wave fluctuations, i.e., either  $(\pi,\pi)$ or $(\pi/3,0)$ here. This is distinctively different from an exciton with a single particle-hole excitation and has been coined  $\pi$-ton in Ref.~\onlinecite{kauch2019b}. 
That means that weak localization corrections in the  $pp$-channel are not the dominating vertex corrections to the optical conductivity, at least not in the parameter regime  studied where the $\pht$-channel is more important. It becomes even more clear for the doped system, where the reduction of conductivity observed within a $pp$-ladder calculation is inverted when applying the full parquet calculation.}


\acknowledgments
We thank Vaclav Jani\v{s}, Gang Li,  and Petra Pudleiner for valuable
discussions, and Josef Kaufmann for assistance with the analytic
continuation.
This work has been supported  by the
Austrian Science Fund (FWF) through  project P 30997.
Calculations have been done in part on the Vienna Scientific Cluster (VSC).


\begin{thebibliography}{42}%
\makeatletter
\providecommand \@ifxundefined [1]{%
 \@ifx{#1\undefined}
}%
\providecommand \@ifnum [1]{%
 \ifnum #1\expandafter \@firstoftwo
 \else \expandafter \@secondoftwo
 \fi
}%
\providecommand \@ifx [1]{%
 \ifx #1\expandafter \@firstoftwo
 \else \expandafter \@secondoftwo
 \fi
}%
\providecommand \natexlab [1]{#1}%
\providecommand \enquote  [1]{``#1''}%
\providecommand \bibnamefont  [1]{#1}%
\providecommand \bibfnamefont [1]{#1}%
\providecommand \citenamefont [1]{#1}%
\providecommand \href@noop [0]{\@secondoftwo}%
\providecommand \href [0]{\begingroup \@sanitize@url \@href}%
\providecommand \@href[1]{\@@startlink{#1}\@@href}%
\providecommand \@@href[1]{\endgroup#1\@@endlink}%
\providecommand \@sanitize@url [0]{\catcode `\\12\catcode `\$12\catcode
  `\&12\catcode `\#12\catcode `\^12\catcode `\_12\catcode `\%12\relax}%
\providecommand \@@startlink[1]{}%
\providecommand \@@endlink[0]{}%
\providecommand \url  [0]{\begingroup\@sanitize@url \@url }%
\providecommand \@url [1]{\endgroup\@href {#1}{\urlprefix }}%
\providecommand \urlprefix  [0]{URL }%
\providecommand \Eprint [0]{\href }%
\providecommand \doibase [0]{http://dx.doi.org/}%
\providecommand \selectlanguage [0]{\@gobble}%
\providecommand \bibinfo  [0]{\@secondoftwo}%
\providecommand \bibfield  [0]{\@secondoftwo}%
\providecommand \translation [1]{[#1]}%
\providecommand \BibitemOpen [0]{}%
\providecommand \bibitemStop [0]{}%
\providecommand \bibitemNoStop [0]{.\EOS\space}%
\providecommand \EOS [0]{\spacefactor3000\relax}%
\providecommand \BibitemShut  [1]{\csname bibitem#1\endcsname}%
\let\auto@bib@innerbib\@empty
\bibitem [{\citenamefont {Falicov}\ and\ \citenamefont
  {Kimball}(1969)}]{Falicov1969}%
  \BibitemOpen
  \bibfield  {author} {\bibinfo {author} {\bibfnamefont {L.~M.}\ \bibnamefont
  {Falicov}}\ and\ \bibinfo {author} {\bibfnamefont {J.~C.}\ \bibnamefont
  {Kimball}},\ }\href {\doibase 10.1103/PhysRevLett.22.997} {\bibfield
  {journal} {\bibinfo  {journal} {Phys. Rev. Lett.}\ }\textbf {\bibinfo
  {volume} {22}},\ \bibinfo {pages} {997} (\bibinfo {year} {1969})}\BibitemShut
  {NoStop}%
\bibitem [{\citenamefont {Hubbard}(1963)}]{Hubbard1963}%
  \BibitemOpen
  \bibfield  {author} {\bibinfo {author} {\bibfnamefont {J.}~\bibnamefont
  {Hubbard}},\ }\href {\doibase 10.1098/rspa.1963.0204} {\bibfield  {journal}
  {\bibinfo  {journal} {Proceedings of the Royal Society of London. Series A,
  Mathematical and Physical Sciences}\ }\textbf {\bibinfo {volume} {276}},\
  \bibinfo {pages} {238} (\bibinfo {year} {1963})}\BibitemShut {NoStop}%
\bibitem [{\citenamefont {Kennedy}\ and\ \citenamefont
  {Lieb}(1986)}]{Kennedy1986}%
  \BibitemOpen
  \bibfield  {author} {\bibinfo {author} {\bibfnamefont {T.}~\bibnamefont
  {Kennedy}}\ and\ \bibinfo {author} {\bibfnamefont {E.~H.}\ \bibnamefont
  {Lieb}},\ }\href {\doibase 10.1016/0378-4371(86)90188-3} {\bibfield
  {journal} {\bibinfo  {journal} {Physica A}\ }\textbf {\bibinfo {volume}
  {138}},\ \bibinfo {pages} {320} (\bibinfo {year} {1986})}\BibitemShut
  {NoStop}%
\bibitem [{\citenamefont {Brandt}\ and\ \citenamefont
  {Schmidt}(1986)}]{Brandt1986}%
  \BibitemOpen
  \bibfield  {author} {\bibinfo {author} {\bibfnamefont {U.}~\bibnamefont
  {Brandt}}\ and\ \bibinfo {author} {\bibfnamefont {R.}~\bibnamefont
  {Schmidt}},\ }\href {\doibase 10.1007/BF01312577} {\bibfield  {journal}
  {\bibinfo  {journal} {Zeitschrift f{\"u}r Phys. B Condens. Matter}\ }\textbf
  {\bibinfo {volume} {63}},\ \bibinfo {pages} {45} (\bibinfo {year}
  {1986})}\BibitemShut {NoStop}%
\bibitem [{\citenamefont {Metzner}\ and\ \citenamefont
  {Vollhardt}(1989)}]{Metzner1989}%
  \BibitemOpen
  \bibfield  {author} {\bibinfo {author} {\bibfnamefont {W.}~\bibnamefont
  {Metzner}}\ and\ \bibinfo {author} {\bibfnamefont {D.}~\bibnamefont
  {Vollhardt}},\ }\href {\doibase 10.1103/PhysRevLett.62.324} {\bibfield
  {journal} {\bibinfo  {journal} {Phys. Rev. Lett.}\ }\textbf {\bibinfo
  {volume} {62}},\ \bibinfo {pages} {324} (\bibinfo {year} {1989})}\BibitemShut
  {NoStop}%
\bibitem [{\citenamefont {Georges}\ and\ \citenamefont
  {Kotliar}(1992)}]{Georges1992a}%
  \BibitemOpen
  \bibfield  {author} {\bibinfo {author} {\bibfnamefont {A.}~\bibnamefont
  {Georges}}\ and\ \bibinfo {author} {\bibfnamefont {G.}~\bibnamefont
  {Kotliar}},\ }\href {\doibase 10.1103/PhysRevB.45.6479} {\bibfield  {journal}
  {\bibinfo  {journal} {Phys. Rev. B}\ }\textbf {\bibinfo {volume} {45}},\
  \bibinfo {pages} {6479} (\bibinfo {year} {1992})}\BibitemShut {NoStop}%
\bibitem [{\citenamefont {Jani{\v{s}}}(1991)}]{Janis1991}%
  \BibitemOpen
  \bibfield  {author} {\bibinfo {author} {\bibfnamefont {V.}~\bibnamefont
  {Jani{\v{s}}}},\ }\href {\doibase 10.1007/BF01309423} {\bibfield  {journal}
  {\bibinfo  {journal} {Zeitschrift f{\"u}r Physik B Condensed Matter}\
  }\textbf {\bibinfo {volume} {83}},\ \bibinfo {pages} {227} (\bibinfo {year}
  {1991})}\BibitemShut {NoStop}%
\bibitem [{\citenamefont {Freericks}\ and\ \citenamefont
  {Zlati\ifmmode~\acute{c}\else \'{c}\fi{}}(2003)}]{Freericks2003}%
  \BibitemOpen
  \bibfield  {author} {\bibinfo {author} {\bibfnamefont {J.~K.}\ \bibnamefont
  {Freericks}}\ and\ \bibinfo {author} {\bibfnamefont {V.}~\bibnamefont
  {Zlati\ifmmode~\acute{c}\else \'{c}\fi{}}},\ }\href {\doibase
  10.1103/RevModPhys.75.1333} {\bibfield  {journal} {\bibinfo  {journal} {Rev.
  Mod. Phys.}\ }\textbf {\bibinfo {volume} {75}},\ \bibinfo {pages} {1333}
  (\bibinfo {year} {2003})}\BibitemShut {NoStop}%
\bibitem [{Note1()}]{Note1}%
  \BibitemOpen
  \bibinfo {note} {For the extended FKM cf.~Refs.~\protect \rev@citealpnum
  {Lemanski2017} and \protect \rev@citealpnum {Kapcia2019}}\BibitemShut
  {NoStop}%
\bibitem [{\citenamefont {Maier}\ \emph {et~al.}(2005)\citenamefont {Maier},
  \citenamefont {Jarrell}, \citenamefont {Pruschke},\ and\ \citenamefont
  {Hettler}}]{Maier2005}%
  \BibitemOpen
  \bibfield  {author} {\bibinfo {author} {\bibfnamefont {T.}~\bibnamefont
  {Maier}}, \bibinfo {author} {\bibfnamefont {M.}~\bibnamefont {Jarrell}},
  \bibinfo {author} {\bibfnamefont {T.}~\bibnamefont {Pruschke}}, \ and\
  \bibinfo {author} {\bibfnamefont {M.~H.}\ \bibnamefont {Hettler}},\ }\href
  {\doibase 10.1103/RevModPhys.77.1027} {\bibfield  {journal} {\bibinfo
  {journal} {Rev. Mod. Phys.}\ }\textbf {\bibinfo {volume} {77}},\ \bibinfo
  {pages} {1027} (\bibinfo {year} {2005})}\BibitemShut {NoStop}%
\bibitem [{\citenamefont {Toschi}\ \emph {et~al.}(2007)\citenamefont {Toschi},
  \citenamefont {Katanin},\ and\ \citenamefont {Held}}]{Toschi2007}%
  \BibitemOpen
  \bibfield  {author} {\bibinfo {author} {\bibfnamefont {A.}~\bibnamefont
  {Toschi}}, \bibinfo {author} {\bibfnamefont {A.~A.}\ \bibnamefont {Katanin}},
  \ and\ \bibinfo {author} {\bibfnamefont {K.}~\bibnamefont {Held}},\ }\href
  {\doibase 10.1103/PhysRevB.75.045118} {\bibfield  {journal} {\bibinfo
  {journal} {Phys Rev. B}\ }\textbf {\bibinfo {volume} {75}},\ \bibinfo {pages}
  {045118} (\bibinfo {year} {2007})}\BibitemShut {NoStop}%
\bibitem [{\citenamefont {Katanin}\ \emph {et~al.}(2009)\citenamefont
  {Katanin}, \citenamefont {Toschi},\ and\ \citenamefont {Held}}]{Katanin2009}%
  \BibitemOpen
  \bibfield  {author} {\bibinfo {author} {\bibfnamefont {A.~A.}\ \bibnamefont
  {Katanin}}, \bibinfo {author} {\bibfnamefont {A.}~\bibnamefont {Toschi}}, \
  and\ \bibinfo {author} {\bibfnamefont {K.}~\bibnamefont {Held}},\ }\href
  {\doibase 10.1103/PhysRevB.80.075104} {\bibfield  {journal} {\bibinfo
  {journal} {Phys. Rev. B}\ }\textbf {\bibinfo {volume} {80}},\ \bibinfo
  {pages} {075104} (\bibinfo {year} {2009})}\BibitemShut {NoStop}%
\bibitem [{\citenamefont {Kusunose}(2006)}]{Kusunose2006}%
  \BibitemOpen
  \bibfield  {author} {\bibinfo {author} {\bibfnamefont {H.}~\bibnamefont
  {Kusunose}},\ }\href {\doibase 10.1143/JPSJ.75.054713} {\bibfield  {journal}
  {\bibinfo  {journal} {J. Phys. Soc. Jpn.}\ }\textbf {\bibinfo {volume}
  {75}},\ \bibinfo {pages} {054713} (\bibinfo {year} {2006})}\BibitemShut
  {NoStop}%
\bibitem [{\citenamefont {Rubtsov}\ \emph {et~al.}(2008)\citenamefont
  {Rubtsov}, \citenamefont {Katsnelson},\ and\ \citenamefont
  {Lichtenstein}}]{Rubtsov2008}%
  \BibitemOpen
  \bibfield  {author} {\bibinfo {author} {\bibfnamefont {A.~N.}\ \bibnamefont
  {Rubtsov}}, \bibinfo {author} {\bibfnamefont {M.~I.}\ \bibnamefont
  {Katsnelson}}, \ and\ \bibinfo {author} {\bibfnamefont {A.~I.}\ \bibnamefont
  {Lichtenstein}},\ }\href {\doibase 10.1103/PhysRevB.77.033101} {\bibfield
  {journal} {\bibinfo  {journal} {Phys. Rev. B}\ }\textbf {\bibinfo {volume}
  {77}},\ \bibinfo {pages} {033101} (\bibinfo {year} {2008})}\BibitemShut
  {NoStop}%
\bibitem [{\citenamefont {Jani\v{s}}(2001)}]{Janis2001}%
  \BibitemOpen
  \bibfield  {author} {\bibinfo {author} {\bibfnamefont {V.}~\bibnamefont
  {Jani\v{s}}},\ }\href {\doibase 10.1103/PhysRevB.64.115115} {\bibfield
  {journal} {\bibinfo  {journal} {Phys. Rev. B}\ }\textbf {\bibinfo {volume}
  {64}},\ \bibinfo {pages} {115115} (\bibinfo {year} {2001})}\BibitemShut
  {NoStop}%
\bibitem [{\citenamefont {Rohringer}\ \emph {et~al.}(2018)\citenamefont
  {Rohringer}, \citenamefont {Hafermann}, \citenamefont {Toschi}, \citenamefont
  {Katanin}, \citenamefont {Antipov}, \citenamefont {Katsnelson}, \citenamefont
  {Lichtenstein}, \citenamefont {Rubtsov},\ and\ \citenamefont
  {Held}}]{RMPvertex}%
  \BibitemOpen
  \bibfield  {author} {\bibinfo {author} {\bibfnamefont {G.}~\bibnamefont
  {Rohringer}}, \bibinfo {author} {\bibfnamefont {H.}~\bibnamefont
  {Hafermann}}, \bibinfo {author} {\bibfnamefont {A.}~\bibnamefont {Toschi}},
  \bibinfo {author} {\bibfnamefont {A.~A.}\ \bibnamefont {Katanin}}, \bibinfo
  {author} {\bibfnamefont {A.~E.}\ \bibnamefont {Antipov}}, \bibinfo {author}
  {\bibfnamefont {M.~I.}\ \bibnamefont {Katsnelson}}, \bibinfo {author}
  {\bibfnamefont {A.~I.}\ \bibnamefont {Lichtenstein}}, \bibinfo {author}
  {\bibfnamefont {A.~N.}\ \bibnamefont {Rubtsov}}, \ and\ \bibinfo {author}
  {\bibfnamefont {K.}~\bibnamefont {Held}},\ }\href {\doibase
  10.1103/RevModPhys.90.025003} {\bibfield  {journal} {\bibinfo  {journal}
  {Rev. Mod. Phys.}\ }\textbf {\bibinfo {volume} {90}},\ \bibinfo {pages}
  {025003} (\bibinfo {year} {2018})}\BibitemShut {NoStop}%
\bibitem [{\citenamefont {Antipov}\ \emph {et~al.}(2014)\citenamefont
  {Antipov}, \citenamefont {Gull},\ and\ \citenamefont
  {Kirchner}}]{antipov2014}%
  \BibitemOpen
  \bibfield  {author} {\bibinfo {author} {\bibfnamefont {A.~E.}\ \bibnamefont
  {Antipov}}, \bibinfo {author} {\bibfnamefont {E.}~\bibnamefont {Gull}}, \
  and\ \bibinfo {author} {\bibfnamefont {S.}~\bibnamefont {Kirchner}},\ }\href
  {\doibase 10.1103/PhysRevLett.112.226401} {\bibfield  {journal} {\bibinfo
  {journal} {Phys. Rev. Lett.}\ }\textbf {\bibinfo {volume} {112}},\ \bibinfo
  {pages} {226401} (\bibinfo {year} {2014})}\BibitemShut {NoStop}%
\bibitem [{\citenamefont {Ribic}\ \emph {et~al.}(2016)\citenamefont {Ribic},
  \citenamefont {Rohringer},\ and\ \citenamefont {Held}}]{ribic2016}%
  \BibitemOpen
  \bibfield  {author} {\bibinfo {author} {\bibfnamefont {T.}~\bibnamefont
  {Ribic}}, \bibinfo {author} {\bibfnamefont {G.}~\bibnamefont {Rohringer}}, \
  and\ \bibinfo {author} {\bibfnamefont {K.}~\bibnamefont {Held}},\ }\href
  {\doibase 10.1103/PhysRevB.93.195105} {\bibfield  {journal} {\bibinfo
  {journal} {Phys. Rev. B}\ }\textbf {\bibinfo {volume} {93}},\ \bibinfo
  {pages} {195105} (\bibinfo {year} {2016})}\BibitemShut {NoStop}%
\bibitem [{\citenamefont {Yang}\ \emph {et~al.}(2014)\citenamefont {Yang},
  \citenamefont {Haase}, \citenamefont {Terletska}, \citenamefont {Meng},
  \citenamefont {Pruschke}, \citenamefont {Moreno},\ and\ \citenamefont
  {Jarrell}}]{yang2014}%
  \BibitemOpen
  \bibfield  {author} {\bibinfo {author} {\bibfnamefont {S.-X.}\ \bibnamefont
  {Yang}}, \bibinfo {author} {\bibfnamefont {P.}~\bibnamefont {Haase}},
  \bibinfo {author} {\bibfnamefont {H.}~\bibnamefont {Terletska}}, \bibinfo
  {author} {\bibfnamefont {Z.~Y.}\ \bibnamefont {Meng}}, \bibinfo {author}
  {\bibfnamefont {T.}~\bibnamefont {Pruschke}}, \bibinfo {author}
  {\bibfnamefont {J.}~\bibnamefont {Moreno}}, \ and\ \bibinfo {author}
  {\bibfnamefont {M.}~\bibnamefont {Jarrell}},\ }\href {\doibase
  10.1103/PhysRevB.89.195116} {\bibfield  {journal} {\bibinfo  {journal} {Phys.
  Rev. B}\ }\textbf {\bibinfo {volume} {89}},\ \bibinfo {pages} {195116}
  (\bibinfo {year} {2014})}\BibitemShut {NoStop}%
\bibitem [{\citenamefont {Altshuler}\ and\ \citenamefont
  {Aronov}(1985)}]{Altshuler1985}%
  \BibitemOpen
  \bibfield  {author} {\bibinfo {author} {\bibfnamefont {B.~L.}\ \bibnamefont
  {Altshuler}}\ and\ \bibinfo {author} {\bibfnamefont {A.~G.}\ \bibnamefont
  {Aronov}},\ }\href@noop {} {\emph {\bibinfo {title} {Electron-Electron
  interaction in disordered conductors}}},\ edited by\ \bibinfo {editor}
  {\bibfnamefont {A.~I.}\ \bibnamefont {Efros}}\ and\ \bibinfo {editor}
  {\bibfnamefont {M.}~\bibnamefont {Pollak}}\ (\bibinfo  {publisher} {Elsevier
  Science Publisher},\ \bibinfo {year} {1985})\BibitemShut {NoStop}%
\bibitem [{\citenamefont {Antipov}\ \emph {et~al.}(2016)\citenamefont
  {Antipov}, \citenamefont {Javanmard}, \citenamefont {Ribeiro},\ and\
  \citenamefont {Kirchner}}]{Antipov2016}%
  \BibitemOpen
  \bibfield  {author} {\bibinfo {author} {\bibfnamefont {A.~E.}\ \bibnamefont
  {Antipov}}, \bibinfo {author} {\bibfnamefont {Y.}~\bibnamefont {Javanmard}},
  \bibinfo {author} {\bibfnamefont {P.}~\bibnamefont {Ribeiro}}, \ and\
  \bibinfo {author} {\bibfnamefont {S.}~\bibnamefont {Kirchner}},\ }\href
  {\doibase 10.1103/PhysRevLett.117.146601} {\bibfield  {journal} {\bibinfo
  {journal} {Phys. Rev. Lett.}\ }\textbf {\bibinfo {volume} {117}},\ \bibinfo
  {pages} {146601} (\bibinfo {year} {2016})}\BibitemShut {NoStop}%
\bibitem [{\citenamefont {\ifmmode~\check{Z}\else \v{Z}\fi{}onda}\ and\
  \citenamefont {Thoss}(2019)}]{Zonda2019}%
  \BibitemOpen
  \bibfield  {author} {\bibinfo {author} {\bibfnamefont {M.}~\bibnamefont
  {\ifmmode~\check{Z}\else \v{Z}\fi{}onda}}\ and\ \bibinfo {author}
  {\bibfnamefont {M.}~\bibnamefont {Thoss}},\ }\href {\doibase
  10.1103/PhysRevB.99.155157} {\bibfield  {journal} {\bibinfo  {journal} {Phys.
  Rev. B}\ }\textbf {\bibinfo {volume} {99}},\ \bibinfo {pages} {155157}
  (\bibinfo {year} {2019})}\BibitemShut {NoStop}%
\bibitem [{Note2()}]{Note2}%
  \BibitemOpen
  \bibinfo {note} {A diagram is two-particle reducible if it can be separated
  into two diagrams by cutting two Green's function lines.}\BibitemShut {Stop}%
\bibitem [{\citenamefont {{Kauch}}\ \emph {et~al.}(2019)\citenamefont
  {{Kauch}}, \citenamefont {{H{\"o}rbinger}}, \citenamefont {{Li}},\ and\
  \citenamefont {{Held}}}]{Kauch2019}%
  \BibitemOpen
  \bibfield  {author} {\bibinfo {author} {\bibfnamefont {A.}~\bibnamefont
  {{Kauch}}}, \bibinfo {author} {\bibfnamefont {F.}~\bibnamefont
  {{H{\"o}rbinger}}}, \bibinfo {author} {\bibfnamefont {G.}~\bibnamefont
  {{Li}}}, \ and\ \bibinfo {author} {\bibfnamefont {K.}~\bibnamefont
  {{Held}}},\ }\href@noop {} {\bibfield  {journal} {\bibinfo  {journal} {arXiv
  e-prints}\ ,\ \bibinfo {eid} {arXiv:1901.09743}} (\bibinfo {year} {2019})},\
  \Eprint {http://arxiv.org/abs/1901.09743} {arXiv:1901.09743
  [cond-mat.str-el]} \BibitemShut {NoStop}%
\bibitem [{\citenamefont {Rohringer}\ \emph {et~al.}(2012)\citenamefont
  {Rohringer}, \citenamefont {Valli},\ and\ \citenamefont
  {Toschi}}]{rohringer2012}%
  \BibitemOpen
  \bibfield  {author} {\bibinfo {author} {\bibfnamefont {G.}~\bibnamefont
  {Rohringer}}, \bibinfo {author} {\bibfnamefont {A.}~\bibnamefont {Valli}}, \
  and\ \bibinfo {author} {\bibfnamefont {A.}~\bibnamefont {Toschi}},\ }\href
  {\doibase 10.1103/PhysRevB.86.125114} {\bibfield  {journal} {\bibinfo
  {journal} {Phys. Rev. B}\ }\textbf {\bibinfo {volume} {86}},\ \bibinfo
  {pages} {125114} (\bibinfo {year} {2012})}\BibitemShut {NoStop}%
\bibitem [{Note3()}]{Note3}%
  \BibitemOpen
  \bibinfo {note} {Note, that in contrast to Ref.~\protect \rev@citealpnum
  {Rubtsov2008,Rubtsov2009} but in agreement with e.g. Ref~\protect
  \rev@citealpnum {ribic2016}, we use the DF self-energy directly as the
  physical self-energy without transformation formula between both
  self-energies. This is because the transformation formula only holds if
  diagrams from higher order vertices are included.\cite {Katanin2013} For the
  same reason, we directly use the full impurity vertex, without using the same
  transformation factors as for the self-energy to arrive the DF interaction,
  cf. Ref.~\protect \rev@citealpnum {Rubtsov2009}. In our calculations the DF
  self-energy is anyhow rather small so that these transformation formulas
  become just the identity up to very small corrections.}\BibitemShut {Stop}%
\bibitem [{\citenamefont {Geffroy}\ \emph {et~al.}(2019)\citenamefont
  {Geffroy}, \citenamefont {Kaufmann}, \citenamefont {Hariki}, \citenamefont
  {Gunacker}, \citenamefont {Hausoel},\ and\ \citenamefont
  {Kune\ifmmode~\check{s}\else \v{s}\fi{}}}]{Kaufmann2018}%
  \BibitemOpen
  \bibfield  {author} {\bibinfo {author} {\bibfnamefont {D.}~\bibnamefont
  {Geffroy}}, \bibinfo {author} {\bibfnamefont {J.}~\bibnamefont {Kaufmann}},
  \bibinfo {author} {\bibfnamefont {A.}~\bibnamefont {Hariki}}, \bibinfo
  {author} {\bibfnamefont {P.}~\bibnamefont {Gunacker}}, \bibinfo {author}
  {\bibfnamefont {A.}~\bibnamefont {Hausoel}}, \ and\ \bibinfo {author}
  {\bibfnamefont {J.}~\bibnamefont {Kune\ifmmode~\check{s}\else \v{s}\fi{}}},\
  }\href {\doibase 10.1103/PhysRevLett.122.127601} {\bibfield  {journal}
  {\bibinfo  {journal} {Phys. Rev. Lett.}\ }\textbf {\bibinfo {volume} {122}},\
  \bibinfo {pages} {127601} (\bibinfo {year} {2019})}\BibitemShut {NoStop}%
\bibitem [{\citenamefont {Astretsov}\ \emph {et~al.}(2019)\citenamefont
  {Astretsov}, \citenamefont {Rohringer},\ and\ \citenamefont
  {Rubtsov}}]{Astretsov2019}%
  \BibitemOpen
  \bibfield  {author} {\bibinfo {author} {\bibfnamefont {G.~V.}\ \bibnamefont
  {Astretsov}}, \bibinfo {author} {\bibfnamefont {G.}~\bibnamefont
  {Rohringer}}, \ and\ \bibinfo {author} {\bibfnamefont {A.~N.}\ \bibnamefont
  {Rubtsov}},\ }\href@noop {} {\bibfield  {journal} {\bibinfo  {journal} {arXiv
  e-prints}\ } (\bibinfo {year} {2019})},\ \Eprint
  {http://arxiv.org/abs/1910.03525} {arXiv:1910.03525 [cond-mat.str-el]}
  \BibitemShut {NoStop}%
\bibitem [{\citenamefont {{Eckhardt}}\ \emph {et~al.}(2019)\citenamefont
  {{Eckhardt}}, \citenamefont {{Honerkamp}}, \citenamefont {{Held}},\ and\
  \citenamefont {{Kauch}}}]{TUPS}%
  \BibitemOpen
  \bibfield  {author} {\bibinfo {author} {\bibfnamefont {C.~J.}\ \bibnamefont
  {{Eckhardt}}}, \bibinfo {author} {\bibfnamefont {C.}~\bibnamefont
  {{Honerkamp}}}, \bibinfo {author} {\bibfnamefont {K.}~\bibnamefont {{Held}}},
  \ and\ \bibinfo {author} {\bibfnamefont {A.}~\bibnamefont {{Kauch}}},\
  }\href@noop {} {\bibfield  {journal} {\bibinfo  {journal} {arXiv e-prints}\
  ,\ \bibinfo {eid} {arXiv:1912.07469}} (\bibinfo {year} {2019})},\ \Eprint
  {http://arxiv.org/abs/1912.07469} {arXiv:1912.07469 [cond-mat.str-el]}
  \BibitemShut {NoStop}%
\bibitem [{\citenamefont {Abrahams}\ \emph {et~al.}(1979)\citenamefont
  {Abrahams}, \citenamefont {Anderson}, \citenamefont {Licciardello},\ and\
  \citenamefont {Ramakrishnan}}]{Abrahams1979}%
  \BibitemOpen
  \bibfield  {author} {\bibinfo {author} {\bibfnamefont {E.}~\bibnamefont
  {Abrahams}}, \bibinfo {author} {\bibfnamefont {P.~W.}\ \bibnamefont
  {Anderson}}, \bibinfo {author} {\bibfnamefont {D.~C.}\ \bibnamefont
  {Licciardello}}, \ and\ \bibinfo {author} {\bibfnamefont {T.~V.}\
  \bibnamefont {Ramakrishnan}},\ }\href@noop {} {\bibfield  {journal} {\bibinfo
   {journal} {Phys. Rev. Lett.}\ }\textbf {\bibinfo {volume} {42}},\ \bibinfo
  {pages} {673} (\bibinfo {year} {1979})}\BibitemShut {NoStop}%
\bibitem [{Note4()}]{Note4}%
  \BibitemOpen
  \bibinfo {note} {The results presented here are calculated in a $pp$-ladder
  approximation, using equation (\ref {eq:pp_ladder}) to calculate the full
  dual vertex and without updating the propagator of the real electrons to
  emphasize the diagrammatics of the $pp$-ladder without corrections in the
  propagator.}\BibitemShut {Stop}%
\bibitem [{Note5()}]{Note5}%
  \BibitemOpen
  \bibinfo {note} {For similar calculations of vertex corrections in disordered
  systems, cf.~Refs.\protect \rev@citealpnum {Janis2010} and \protect
  \rev@citealpnum {Pokorny2013}.}\BibitemShut {Stop}%
\bibitem [{\citenamefont {Kauch}\ \emph {et~al.}(2020)\citenamefont {Kauch},
  \citenamefont {Pudleiner}, \citenamefont {Astleithner}, \citenamefont
  {Thunstr\"om}, \citenamefont {Ribic},\ and\ \citenamefont
  {Held}}]{kauch2019b}%
  \BibitemOpen
  \bibfield  {author} {\bibinfo {author} {\bibfnamefont {A.}~\bibnamefont
  {Kauch}}, \bibinfo {author} {\bibfnamefont {P.}~\bibnamefont {Pudleiner}},
  \bibinfo {author} {\bibfnamefont {K.}~\bibnamefont {Astleithner}}, \bibinfo
  {author} {\bibfnamefont {P.}~\bibnamefont {Thunstr\"om}}, \bibinfo {author}
  {\bibfnamefont {T.}~\bibnamefont {Ribic}}, \ and\ \bibinfo {author}
  {\bibfnamefont {K.}~\bibnamefont {Held}},\ }\href {\doibase
  10.1103/PhysRevLett.124.047401} {\bibfield  {journal} {\bibinfo  {journal}
  {Phys. Rev. Lett.}\ }\textbf {\bibinfo {volume} {124}},\ \bibinfo {pages}
  {047401} (\bibinfo {year} {2020})}\BibitemShut {NoStop}%
\bibitem [{Note6()}]{Note6}%
  \BibitemOpen
  \bibinfo {note} {The change in the occupancy of the $c$ electrons is also
  reflected in the offset in the real part of the DF self-energy in Fig. \ref
  {fig:sigma_parquet_doped} that differs from the offset in DMFT.}\BibitemShut
  {Stop}%
\bibitem [{cod(2019)}]{code_fkm}%
  \BibitemOpen
  \href {https://github.com/Adepttin/FK-FGA-} {\enquote {\bibinfo {title}
  {https://github.com/adepttin/fk-fga-},}\ } (\bibinfo {year}
  {2019})\BibitemShut {NoStop}%
\bibitem [{\citenamefont {Astleithner}(2019)}]{Astleithner_Thesis}%
  \BibitemOpen
  \bibfield  {author} {\bibinfo {author} {\bibfnamefont {K.}~\bibnamefont
  {Astleithner}},\ }\emph {\bibinfo {title} {Optical conductivity in the
  Falicov-Kimball model: a dual fermion perspective}},\ \href
  {http://repositum.tuwien.ac.at/obvutwhs/content/titleinfo/3601320} {Master's
  thesis},\ \bibinfo  {school} {Vienna University of Technology} (\bibinfo
  {year} {2019})\BibitemShut {NoStop}%
\bibitem [{\citenamefont {Lema\ifmmode~\acute{n}\else \'{n}\fi{}ski}\ \emph
  {et~al.}(2017)\citenamefont {Lema\ifmmode~\acute{n}\else \'{n}\fi{}ski},
  \citenamefont {Kapcia},\ and\ \citenamefont {Robaszkiewicz}}]{Lemanski2017}%
  \BibitemOpen
  \bibfield  {author} {\bibinfo {author} {\bibfnamefont {R.}~\bibnamefont
  {Lema\ifmmode~\acute{n}\else \'{n}\fi{}ski}}, \bibinfo {author}
  {\bibfnamefont {K.~J.}\ \bibnamefont {Kapcia}}, \ and\ \bibinfo {author}
  {\bibfnamefont {S.}~\bibnamefont {Robaszkiewicz}},\ }\href {\doibase
  10.1103/PhysRevB.96.205102} {\bibfield  {journal} {\bibinfo  {journal} {Phys.
  Rev. B}\ }\textbf {\bibinfo {volume} {96}},\ \bibinfo {pages} {205102}
  (\bibinfo {year} {2017})}\BibitemShut {NoStop}%
\bibitem [{\citenamefont {Kapcia}\ \emph {et~al.}(2019)\citenamefont {Kapcia},
  \citenamefont {Lema\ifmmode~\acute{n}\else \'{n}\fi{}ski},\ and\
  \citenamefont {Robaszkiewicz}}]{Kapcia2019}%
  \BibitemOpen
  \bibfield  {author} {\bibinfo {author} {\bibfnamefont {K.~J.}\ \bibnamefont
  {Kapcia}}, \bibinfo {author} {\bibfnamefont {R.}~\bibnamefont
  {Lema\ifmmode~\acute{n}\else \'{n}\fi{}ski}}, \ and\ \bibinfo {author}
  {\bibfnamefont {S.}~\bibnamefont {Robaszkiewicz}},\ }\href {\doibase
  10.1103/PhysRevB.99.245143} {\bibfield  {journal} {\bibinfo  {journal} {Phys.
  Rev. B}\ }\textbf {\bibinfo {volume} {99}},\ \bibinfo {pages} {245143}
  (\bibinfo {year} {2019})}\BibitemShut {NoStop}%
\bibitem [{\citenamefont {Rubtsov}\ \emph {et~al.}(2009)\citenamefont
  {Rubtsov}, \citenamefont {Katsnelson}, \citenamefont {Lichtenstein},\ and\
  \citenamefont {Georges}}]{Rubtsov2009}%
  \BibitemOpen
  \bibfield  {author} {\bibinfo {author} {\bibfnamefont {A.~N.}\ \bibnamefont
  {Rubtsov}}, \bibinfo {author} {\bibfnamefont {M.~I.}\ \bibnamefont
  {Katsnelson}}, \bibinfo {author} {\bibfnamefont {A.~I.}\ \bibnamefont
  {Lichtenstein}}, \ and\ \bibinfo {author} {\bibfnamefont {A.}~\bibnamefont
  {Georges}},\ }\href {\doibase 10.1103/PhysRevB.79.045133} {\bibfield
  {journal} {\bibinfo  {journal} {Phys. Rev. B}\ }\textbf {\bibinfo {volume}
  {79}},\ \bibinfo {pages} {045133} (\bibinfo {year} {2009})}\BibitemShut
  {NoStop}%
\bibitem [{\citenamefont {Katanin}(2013)}]{Katanin2013}%
  \BibitemOpen
  \bibfield  {author} {\bibinfo {author} {\bibfnamefont {A.~A.}\ \bibnamefont
  {Katanin}},\ }\href {http://stacks.iop.org/1751-8121/46/i=4/a=045002}
  {\bibfield  {journal} {\bibinfo  {journal} {J. Phys. A: Math. Theor.}\
  }\textbf {\bibinfo {volume} {46}},\ \bibinfo {pages} {045002} (\bibinfo
  {year} {2013})}\BibitemShut {NoStop}%
\bibitem [{\citenamefont {Jani{\v{s}}}\ and\ \citenamefont
  {Pokorn\'y}(2010)}]{Janis2010}%
  \BibitemOpen
  \bibfield  {author} {\bibinfo {author} {\bibfnamefont {V.}~\bibnamefont
  {Jani{\v{s}}}}\ and\ \bibinfo {author} {\bibfnamefont {V.}~\bibnamefont
  {Pokorn\'y}},\ }\href {\doibase 10.1103/PhysRevB.81.165103} {\bibfield
  {journal} {\bibinfo  {journal} {Phys. Rev. B}\ }\textbf {\bibinfo {volume}
  {81}},\ \bibinfo {pages} {165103} (\bibinfo {year} {2010})}\BibitemShut
  {NoStop}%
\bibitem [{\citenamefont {Pokorny}\ and\ \citenamefont
  {Janis}(2013)}]{Pokorny2013}%
  \BibitemOpen
  \bibfield  {author} {\bibinfo {author} {\bibfnamefont {V.}~\bibnamefont
  {Pokorny}}\ and\ \bibinfo {author} {\bibfnamefont {V.}~\bibnamefont
  {Janis}},\ }\href {\doibase 10.1088/0953-8984/25/17/175502} {\bibfield
  {journal} {\bibinfo  {journal} {J. Phys. Condens. Matter}\ }\textbf {\bibinfo
  {volume} {25}},\ \bibinfo {pages} {175502} (\bibinfo {year}
  {2013})}\BibitemShut {NoStop}%
\end{thebibliography}

%

\end{document}